\documentclass[aps,twocolumn,superscriptaddress,floatfix]{revtex4-1}
\usepackage{bbm} 
\usepackage{amsmath} 
\usepackage{amssymb} 
\usepackage{physics} 
\usepackage[bookmarks=true]{hyperref} 
\usepackage[english]{babel} 
\usepackage[normalem]{ulem} 
\usepackage{xfrac}
\usepackage{graphicx}
\usepackage[]{xcolor} 

\allowdisplaybreaks

\begin{document}

\title{Casimir Physics beyond the Proximity Force Approximation: \linebreak  The Derivative Expansion}

\author{C\'esar D. Fosco} 
\affiliation{Centro At\'omico Bariloche, Instituto Balseiro,
Comisi\'on Nacional de Energ\'\i a At\'omica, 
8400 Bariloche, Argentina} 
\author{Fernando C. Lombardo}
\affiliation{Departamento de Física, Facultad de Ciencias Exactas y Naturales, Universidad de Buenos Aires, Buenos Aires, Argentina}
\affiliation{ Instituto de Física de Buenos Aires (IFIBA), CONICET, Universidad de Buenos Aires, Argentina}
\author{Fransisc D. Mazzitelli}
\affiliation{Centro At\'omico Bariloche, Instituto Balseiro,
Comisi\'on Nacional de Energ\'\i a At\'omica, 
8400 Bariloche, Argentina}

\begin{abstract} 
We review the derivative expansion (DE) method in Casimir physics, an approach which extends the proximity force approximation (PFA). After introducing and motivating the DE in contexts other than the Casimir effect, we present different examples which correspond to that realm. We focus on different particular geometries, boundary conditions, types of fields, and quantum and thermal fluctuations. Besides providing various examples where the method can be applied, we discuss a concrete example for which the DE cannot be applied; namely, the case of perfect Neumann conditions in 2 + 1 dimensions. By the same example, we show how a more realistic type of boundary condition circumvents the problem. We also comment on the application of the DE to the Casimir–Polder interaction which provides a broader perspective on particle–surface interactions.
\end{abstract}

\maketitle
\section{Introduction}
Casimir forces are one of the most intriguing macroscopic manifestations of quantum fluctuations in 
Nature. Their existence, first realized in the specific context of the interaction between the
quantum electromagnetic (EM) field and the boundaries of two neutral bodies, manifests itself 
as an attractive force between them. That force depends, in an intricate manner, on the shape 
and  EM properties of the objects. Since the discovery of this effect  by Hendrik Casimir 75 years ago \cite{Casimir48} this, and closely related phenomena, have been subjected to  intense theoretical and experimental  research \cite{Milonnibook,Miltonbook,Bordagbook,Dalvitbook}. The outcome of that work has not just revealed fundamental aspects of quantum field theory,  but also subtle aspects of 
the models used to describe the EM  properties of material bodies.  Besides, it has become increasingly
clear that this research  has potential applications to nanotechnology.

Theoretical and experimental reasons have called for the calculation of the Casimir energies and forces 
for different geometries and materials \cite{Mostepanenko2021}, and with an ever increasing accuracy.  The simplicity of  the theoretical predictions when two parallel plates are involved, corresponds to a  difficult experimental  setup, due to alignment problems (in spite of this, the Casimir force for this geometry has been measured at the $10\%$ accuracy level \cite{Carugno}). Equivalently, geometries which are 
more convenient from the  experimental point of view, and allow for higher precision measurements, 
lead to more involved theoretical calculations. Such is the case of a cylinder facing a plane \cite{onofrio}, or a sphere facing a plate,  which is free from the alluded alignment problems \cite{precisionera1,precisionera2,precisionera3,precisionera4,precisionera5,precisionera6,precisionera7}.

From a theoretical standpoint, finding the dependence of the Casimir energies and forces on 
the geometry of the objects, poses an interesting challenge. Indeed, even when evaluating the self-energies 
which result from the coupling on an object to the vacuum field fluctuations, results may be rather 
non-intuitive; as in the case of a single spherical surface \cite{Boyer68}.

For a long time, calculations attempting to find analytical results for the Casimir and related 
interactions had been restricted to using the so called proximity force approximation (PFA). 
In this approach, the interaction energies and the resulting forces are computed approximating 
the geometry by a collection of parallel plates and then adding up the contributions obtained 
for this  approximate geometry.  This procedure was presumed to work  well enough, at least  for 
smooth surfaces when they are sufficiently close to each other; in more precise terms:  when the
curvature radii of the surfaces $R_i$ are much larger than the distance $d$  between them.   
Indeed, this is the main content of the Derjaguin approximation (DA), developed by Boris Derjaguin
in the 1930s~\cite{Derjaguin1,Derjaguin2,Derjaguin3} , which is pivotal in the study of surface interactions, especially in 
the context of colloidal particles and biological cells. This approach has significant implications in understanding 
colloidal stability, adhesion, and thin film formation.

It is worth introducing some essentials of the DA, in particular, of the geometrical aspects involved. 
Assuming the interaction energy per unit area between two parallel planes at a distance $h$ is known, 
and given by $E_{\shortparallel}(h)$, the DA yields an expression for the interaction energy between two 
curved surfaces, $U_{\rm DA}$~\cite{Milonnibook,Bordagbook,Derjaguin1,Derjaguin2,Derjaguin3,Intermolecular}. Indeed,
\begin{equation}
U_{\rm DA}(a)=2\pi R_{\rm eff}\int_a^\infty E_{\shortparallel}(h)dh\, ,
\label{DA}
\end{equation}  
where  $a$ denotes the distance between the surfaces, $R_1$ and $R_2$ are their curvature 
radii (at the point of closest distance), while $R_{\rm eff}=R_1R_2/(R_1+R_2)$.  
It is rather straightforward to implement the approximation at the level of the force $f_{\rm DA}$ between
surfaces:
\begin{equation}
f_{\rm DA}(a)=2\pi R_{\rm eff} E_{\shortparallel}(a)\, .
\label{DAforce}
\end{equation}  
This approximation is usually derived from   a quite reasonable  assumption, namely, 
that the interaction energy  can be approximated  by means of the PFA expression:
\begin{equation}
U_{\rm PFA}=\int \, dS\, E_{\shortparallel}\; .
\label{PFA}
\end{equation}
Here,  the surface integration may be performed over  one of the participating surfaces, but it could also be  
over an imaginary, ``interpolating surface'',  which lies between them. 
The DA is  obtained from the expression above, by approximating the surfaces  by (portions of) the osculating spheres (with radii $R_1$ and $R_2$) at the point of closest approach.

Based on this hypothesis, on dimensional grounds one can expect the corrections to the PFA
to be of order ${\mathcal O}(a/R_i)$.  Note, however, that since the PFA had not been obtained as the  leading-order
term in a  well-defined expansion, the approximation itself did not provide any quantitative method to
asses the validity of that assumption.  

A need for reliable measure of the accuracy of the results obtained using different methods became
increasingly  crucial,   specially since the development of the ``precision era'' in the measurement of the 
Casimir forces \cite{precisionera1,precisionera2,precisionera3,precisionera4,precisionera5,precisionera6,precisionera7}. It was in this context that the  Derivative Expansion (DE) approach, was first introduced by
us in 2011 \cite{FLM2011}, as a tool  to  asses the validity of the PFA, by putting it in the framework of an expansion, 
and to calculate corrections to the PFA using that very same expansion.
When one realizes that the PFA had previously been proposed in contexts which are rather different
to Casimir physics,  it becomes clear that the improvement on the PFA which represents the DE may and
does have relevance on those realms, regardless of them having an origin in vacuum fluctuations or not. 
Indeed, when one strips off the DE of the particularities of  Casimir physics, one can see the ingredients
that  allowed one to implement it are also found, for example, in  electrostatics, nuclear physics, and 
colloidal surface interactions. 

Here, we present  the essential features of the DE, its derivation, and consider some examples
of its applications.  The review is organized as follows. In Section \ref{sec:nuc} , 
we recall some aspects of the DA which stem from its application to nuclear and colloidal physics. We
start with the DA not just for historical reasons, but also because we believe that
this sheds light on some  geometrical aspects of the approximation, in a rather direct way (like the
relevance of curvature radii and distances).

Then, in Section \ref{sec:ele}, we introduce the DE in  one of its simplest realizations, namely, in the 
context of electrostatics, for a system consisting of two conducting surfaces kept at different potentials \cite{FLMelectro}. 
We first evaluate the PFA in this example, and then introduce the  DE as a method to improve on that approximation.  
In Section \ref{sec:sum}, we introduce a more abstract, and therefore more general, formulation of the 
DE \cite{FLMrev2014}. By putting aside the particular features of an specific interaction, and keeping just the ones
that are common to all of them, we are lead to formulate the problem as follows: the DE is a particular
kind of expansion of a functional having as argument a surface (or surfaces). 
We mean ``functional'' here in its mathematical sense: a function that assigns a number to a function 
or functions.  We elucidate and demonstrate some of the aspects of the DE in this general context; the
purpose of presenting those aspects are not just a matter of consistency or justification, but they also
provide a concrete way of applying and implementing the DE to any example where it is applicable.

Then, in Section \ref{sec:cas}, we focus on the DE in the specific context of the Casimir interaction
between surfaces, for  perfect boundary conditions at zero temperature; i.e., vacuum fluctuations \cite{FLM2011,Bimonte1}.
Then in Section \ref{sec:finite T} we review the extension of those results to the case of finite 
temperatures and real materials \cite{Bimonte2,FLMTdim}. As we shall see, the temperature introduces another scale, which 
affects the form one must adopt for the different terms in the DE.  
Then we comment on an aspect which first manifests itself here: as it happens with any expansion, it is
to be expected to break down for some specific examples, when the hypothesis that justified it are not
satisfied.  We show this for the case of  the Casimir effect with Neumann conditions at finite temperatures \cite{FLMTdim,FLMNeu}.
We also show that the application of the DE to the EM field is free of this problem, if dissipative effects are included in the model describing the media
\cite{FLMTem}. 

The application of the DE to Casimir-Polder forces for atoms near smooth surfaces \cite{BimonteCP} is described in Section \ref{sec:CP}.
Other alternatives to compute Casimir energies beyond PFA \cite{paulo22} are described in Section  \ref{sec:other}. 
Section \ref{sec:conc} contains our conclusions.

\section{Proximity Approximations in Nuclear and Colloidal Physics}\label{sec:nuc}

The introduction of the Derjaguin Approximation (DA) to nuclear physics dates back to the seminal paper \cite{Blocki1}. In this paper, the DA was rediscovered and applied to calculate nuclear interactions, starting with a Derjaguin-like formula for the surface interaction energies.  The approach was based on a crucial "universal function" - a term referring here to the interaction energy between flat surfaces, calculated using a Thomas-Fermi approximation. In spite of the rather different context, the analogy with the approach followed in 
the DA becomes clear when one introduces three surfaces, the physical ones, $\Sigma_L$ and $\Sigma_R$, and the intermediate one $\Sigma$ which one uses to parametrize the interacting ones.  
Then, if the  physical surfaces are sufficiently smooth, the interaction energy should, to a reasonable approximation,  be described by the PFA, in a similar fashion as in Equation (\ref{PFA}). 
To render the assertion above more concrete, we yet again use the function  
\mbox{$h: \Sigma \to {\mathbb R}$},  measuring the distance between   $\Sigma_L$ and $\Sigma_R$ at each  point on $\Sigma$. 
Since $h$  will have  level sets which are, except for a zero measure set, one-dimensional  (closed curves), and the interaction depends just on $h$,  the PFA expression for the interaction energy $U$ may be rendered as a one-dimensional integral:
\begin{equation}
U_{\rm PFA}=\int dh \, J(h) \, E_{\shortparallel}(h)\;,
\label{PFAnucJ}
\end{equation}
where $J(h)dh$ is the infinitesimal area between two level curves on $\Sigma$: the ones between $h$ and $h+dh$, while $E_\shortparallel$ is the universal function. 

We now assume that  $\Sigma$ is a plane,  and that the physical surfaces may be both described by means of just one Monge patch based on $\Sigma$. This surface is then naturally thought of  (in descriptive geometry terms) as the projection plane.  Using  Cartesian coordinates $(x_1,x_2)\equiv {\mathbf x_\shortparallel}$ on $\Sigma$,   assuming (for smooth enough surfaces) that  $J$
may be regarded as constant, and using a  second-order Taylor expansion of $h$  around $a$ (the distance of closest approach):
\begin{equation}
h({\mathbf x_\shortparallel})\simeq
a+\frac{1}{2}\big( \frac{x_1^2}{R_1}+ \frac{x_2^2}{R_2} \big)
\end{equation}
produces, when evaluating the PFA interaction  energy (\ref{PFAnucJ}),   the  DA energy (\ref{DA}). Here, $R_1$ and $R_2$ are the radii of curvature of the surface by  $x_3 = h({\mathbf x_\shortparallel})$ at $x_3=a$.

This result may be improved, even within the spirit of the PFA, by introducing some refinements. Indeed, in~\cite{Blocki2},  a generalization of the PFA has been introduced such that the starting point was Equation (\ref{PFAnucJ}), but now allowing for the surfaces to have larger curvatures, as long as they remained almost parallel locally. 
The main difference that follows from  those weaker assumptions is that, now, the Jacobian $J$ may become a non-trivial function of $h$.  For instance, introducing a linear expansion:
\begin{equation}
J(h)\approx J_0+ J_1 h \;,
\end{equation}
a straightforward calculation shows that the force $f$  becomes:
\begin{equation}\label{eq:correction}
f_{\rm PFA}(a)\,=\, J_0 E_\shortparallel(a)- J_1(a)\int_a^\infty \, dh\,  E_{\shortparallel}(h) \;.
\end{equation}
Note that  the result is  the sum of the DA term plus a  second term  proportional to the derivative of the 
Jacobian with respect to $h$. This is a correction to the  DA obtained from the same starting point we used for the 
DA:  $U_{\rm PFA}$. In other words,  Equation (\ref{eq:correction}) is still determined by the energy
density for parallel plates.  As we shall see,  the DE will introduce corrections that go beyond $E_\shortparallel(a)$. 
The correction will depend on both the geometry and  the nature of the interaction.

We wish to point out that  the lack of knowledge of an exact expression for $E_\shortparallel$ is not specific to 
nuclear physics, but of course it may appear in other applications.  The general PFA approach can nevertheless 
be  introduced;  the accuracy of its predictions will then be limited not just by the fulfillment or not 
of the geometrical  assumptions, but also by the reliability of the expression for $E_\shortparallel$.
 Using different approximations for $E_\shortparallel$ gives as many results for the  PFA.   For a recent review in the
 case of nuclear physics, see Ref.~\cite{RevNuc1,RevNuc2}. 

An apparently unrelated approximation, based on different physical assumptions, was introduced in the 
context of colloidal physics. Let us now see how  it yields a result which agrees with the DA: it is the so 
called Surface Element  Integration (SEI)~\cite{SEI}, or  Surface Integration Approach (SIA)~\cite{SIA}. 
This approach may be introduced as follows:  let us consider a compact object facing the $x_3=0$ plane.
$x_3$ is then the normal coordinate to the plane, pointing towards the  compact object.  With this conventions,
the SEI approximation applied to the interaction energy amounts to the following:
\begin{equation} 
U_{\rm SEI}=-\int_{\rm plane} dx_1 dx_2\, \frac{\hat n\cdot \hat
e_3}{\vert\hat n\cdot\hat e_3\vert} \, E_\shortparallel\;.
\label{SEI}
\end{equation}
Here, $\hat n$ denotes the  outwards pointing unit normal  to each surface element of the  object.
We see that, when the compact object  may be thought of as delimited by just two surfaces, one of them
facing the plane and the other away from it, the SEI  consists of  the difference between the  
PFA energies of  those surfaces.   This (possibly startling) fact is, as we shall see, related to the 
fact that the SEI becomes exact for almost transparent bodies, a situation characterized by the fact that 
the interaction  is the result of  adding all the (volumetric ) pairwise contributions.

In the context of colloidal physics , the SEI method relies heavily  upon the existence
of a pressure on the compact object. The effect of that pressure should be integrated over
the closed surface  surrounding the compact object, in order to find the total force~\cite{SEI}. 
An alternative route to understand the SEI is to showthat Equation (\ref{SEI}) becomes exact when the
interaction between macroscopic bodies is the  superposition of the   interactions for the pair 
potentials of their constituents~\cite{SIA}.  
That  may be interpreted by using a  simple example. Consider  two media, one of them, the left medium $L$,  corresponding to the $x_3 \leq 0$ half-space, while the right medium, $R$, is defined as the region:
\begin{equation}\label{eq:defr}
R \;=\; \{ (x_1,x_2,x_3): \psi_1({\mathbf x_\shortparallel}) \leq x_3 \leq  \psi_2({\mathbf x_\shortparallel}) \}
\,.
\end{equation}
The interaction energy $U$ is a functional of the two functions $\psi_{1,2}$. When the media are diluted, we expect the interaction energy to have the form
\begin{equation}
U[\psi_1, \psi_2] \;=\;  \int d^2 {\mathbf
x_\shortparallel} \left (E_\shortparallel(\psi_1)-E_\shortparallel(\psi_2)\right)\, ,
\label{lin}
\end{equation}
where $E_\shortparallel(a)$ is the interaction energy per unit area,  between  two half-spaces at a distance $a$. 
This formula can be interpreted as follows: to obtain the interaction  energy for the configuration described by $\psi_1$ and $\psi_2$, one must certainly subtract from $E_\shortparallel(\psi_1)$ the contributions from  $x_3>\psi_2$. 
This ``linearity'' is expected to be valid only for dilute media, and in that situation it coincides with
the result obtained using the SEI.   One expects then the  SEI to give an exact result for almost-transparent 
media,  for which  the superposition principle holds true,  and the total interaction  energy is due to the 
sum of all the different pairwise potentials~\cite{SIA}. 
It is worth noting,  at this point, the important fact that the PFA also becomes exact in Casimir physics 
when the media constituting the objects are dilute. Indeed, this has been pointed out  in~\cite{Decca,Miltondilute}.  

The examples just described illustrate the relevance of the DA, and of some of its variants, to different 
areas of physics.  At the same time,  the main drawback is made rather  evident:  in spite of  being based 
on reasonable physical assumptions,  it is  difficult to assess  its validity.  The reason for this difficulty
is that the approximation is uncontrolled, and therefore the estimation of the error incurred is difficult, 
within a self-contained approach. 

The DE provides a systematic method to improve  the PFA, and to compute its next-to-leading-order 
(NTLO) correction in a consistent set up.

\section{Introducing the Derivative Expansion}\label{sec:ele}
\subsection{The PFA in an Electrostatic Example} \label{sec:pfa}

We introduce the PFA, and then the DE, in an example which neatly illustrates the DE main aspects, in the 
context of electrostatics. Here, contrary to what happens when 
dealing with more involved systems, like, say, Van der Waals, nuclear or Casimir forces,  
the physical assumptions and their implementations are more transparent. We follow closely Ref.\cite{FLMelectro}

The set-up we want to describe consists of two perfectly-conducting surfaces, one of them
an infinite grounded  plane and the other a smoothly curved  surface kept at an electrostatic 
potential $V_0$.  We use coordinates such that the plane corresponds to $z=0$  while the smooth
surface is such that it can be described by  a single function, namely, by an equation of the form $z = \psi(x_\shortparallel)$. 
The electrostatic energy contained between surfaces can then be written as follows:
\begin{equation} 
 U=\frac{\epsilon_0}{2}\int d^2{\mathbf x_\shortparallel}\int_0^{\psi({\mathbf x_\shortparallel})} dz\,
 \vert \mathbf E\vert^2\, , 
 \end{equation} 
 where $\epsilon_0$ denotes the permittivity of vacuum. In terms of $U$ and $V_0$, the capacitance 
$C$ of the system is then given  by $C=2U/V_0^2$. 

Let us see  how one implements the PFA in order to calculate $U$ (from which one can extract, for instance, 
an approximate expression for $C$) expecting it to be accurate when the distance between the two surfaces 
is shorter than the curvature radius of the curved conductor.  To that end,  one first  finds and approximation
to the electric field between the conductors, by proceeding as follows: the smooth conductor is regarded as a
set of parallel plates  (Fig.1), in the sense that the electric field ${\mathbf E}$ points along the $z$ direction and has a 
$z$-independent value. The electric field does, however, depend on  ${\mathbf x_\shortparallel}$ since it is assumed to have, for every ${\mathbf x_\shortparallel}$,  the same 
intensity as the electric field due to two (infinite) conducting planes at a distance $\psi({\mathbf x_\shortparallel})$.  Namely,  
${\mathbf E}({\mathbf x}) = - V_0/\psi({\mathbf x_\shortparallel}) \,\mathbf{\hat{z}}$. 
Therefore, the approximated expression for the electrostatic energy becomes:
\begin{equation}
	U_{\rm PFA} =\frac{\epsilon_0 V_0^2}{2}\int d^2{\mathbf x_\shortparallel}\, \frac{1}{\psi({\mathbf x_\shortparallel})}\, .
\label{UPFA}
\end{equation}
It is implicitly understood in the equation above, that the region to integrate is such that the assumption
on the distance and curvature is satisfied. On the contrary, regions such that the assumption is 
not satisfied can be consistently ignored (see the example below).  

It should be evident that Equation (\ref{UPFA}) provides  a rather convenient tool  to obtain  estimates for the electrostatic energy in many relevant situations. Indeed, to illustrate this point we consider a
cylinder of length $L$ and radius $R$ in front of a plane, and denote by $a$ the minimum distance between the two surfaces. The cylinder is not a surface that can be described by a single patch; namely,  one needs at least two functions. However,  in the context of the PFA, it is reasonably to assume that only the half that is closer to the plane should be relevant. Assuming the axis of the cylinder to be along $y$, the function $\psi$  reads: 
\begin{equation}
\psi(x)=a+R\left(1-\sqrt{1-\frac{x^2}{R^2}}\right)\, ,
\label{psicyl}
\end{equation} 
with the variable $x$  assumed to be in the range $-x_M<x<x_M<R$.  Note that for $x_M/R=O(1)<1$
the assumption on the distance and the curvature is satisfied.  It is to be expected that, as long as $R\gg a$ (where  the PFA gives an accurate value of the electrostatic energy), the final result will not depend on $x_M$. This can be readily checked  by
inserting Equation (\ref{psicyl}) into Equation (\ref{UPFA}), computing the integral,  and 
expanding that result for $a\ll R$. Doing this we obtain:
\begin{equation}
U_{\rm PFA}^{\rm cp} \approx\frac{\epsilon_0V_0^2L\pi}{\sqrt{2}}\sqrt{\frac{R}{a}}\, ,
\label{PFAcp}
\end{equation}
which is independent of $x_M$. An immediate consequence of this is that, when the cylinder approaches 
the plane, the electrostatic force behaves as $a^{-3/2}$.

\begin{figure}
\centering
\includegraphics[width=8cm , angle=0]{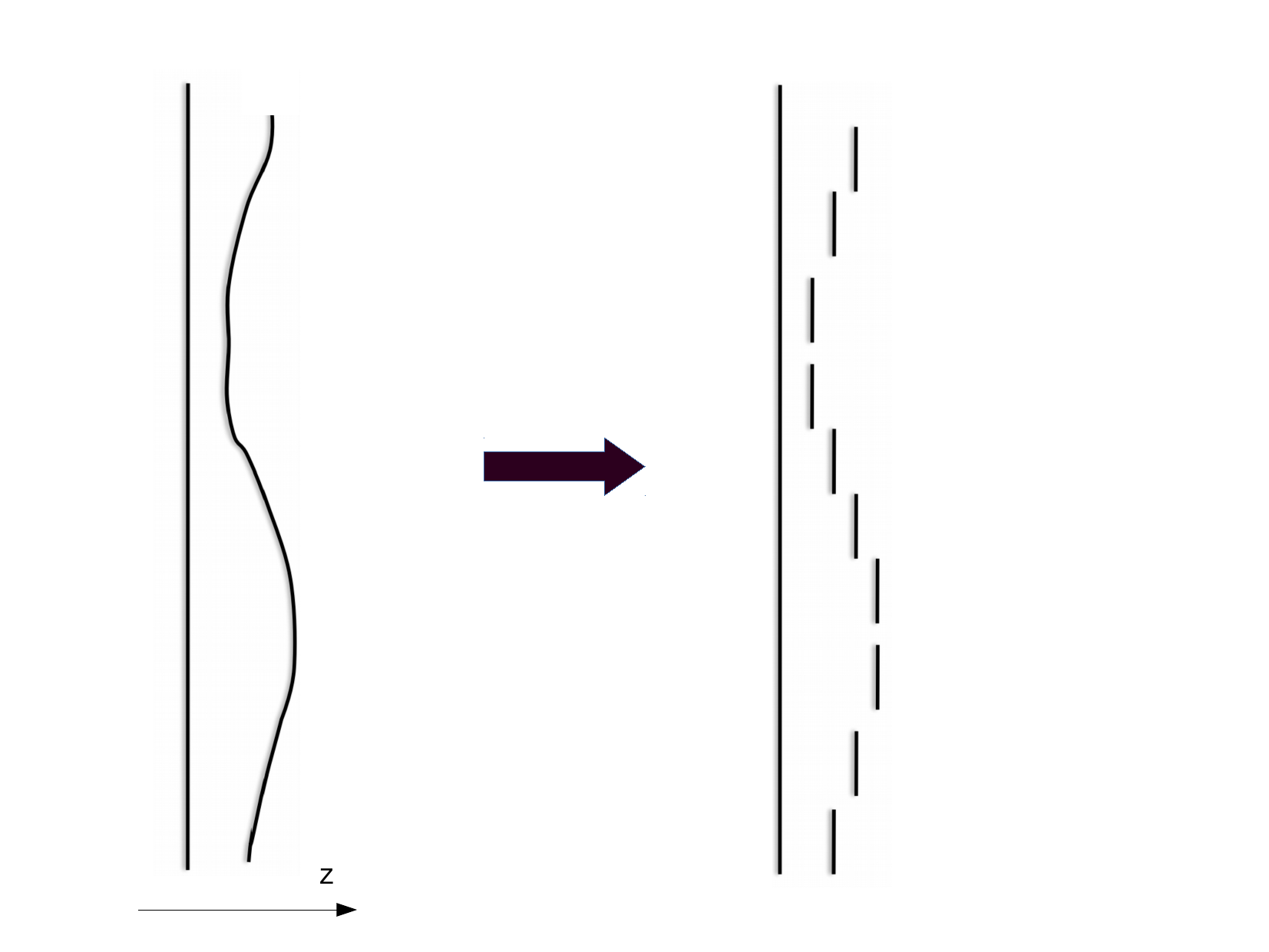}
\caption{\small{In the PFA, the interaction between a smoothly curved surface and a plane is approximated by
that of  a set of parallel plates. For each pair of parallel plates, border effects  are ignored.}} \label{FigPFA}
\end{figure}

Let us check now the accuracy of   $U_{\rm PFA}^{\rm cp}$.  We take advantage of the knowledge
of  the exact expression for the electrostatic interaction energy:
\begin{equation}
 U^{\rm cp} =  \frac{\pi L \epsilon_0 V_0^2}{{\rm arccosh}\left(1 + \frac{a}{R}\right)} \,.
\end{equation}
For $a/R\ll 1$, $U^{\rm cp}$  yields  the PFA result $U^{\rm cp}_{\rm PFA}$ (\ref{PFAcp}).  The relevance of the corrections to the PFA can be estimated by expanding  the exact result, but keeping also the next-to-leading order (NTLO)  when  $a <<R$:
\begin{equation}
 U^{\rm cp} \approx \frac{\epsilon_0V_0^2L\pi}{\sqrt{2}}\sqrt{\frac{R}{a}} \left(1 + \frac{1}{12} \frac{a}{R}\right ) \;.
 \label{cpntlo}
 \end{equation}

We will now introduce the DE. By construction, it should produce the  NTLO result (for this an other surfaces), without resorting to the expansion of any exact expression (the knowledge of which, needless to say, is usually lacking).  

\subsection{Improvement of the PFA Using a Derivative Expansion}\label{ssec:elecde}

We begin by  noting that the electrostatic energy is  a functional
of the function which defines the shape of the surface.  A second  observation 
is that, in principle, there is no reason to assume that  the functional is  local in $\psi$. 
Here, ``local'' means that it contains just one integral  over  ${\mathbf x_\shortparallel}$ of a sum of terms 
involving powers of $\psi({\mathbf x_\shortparallel})$ and  derivatives at $\psi({\mathbf x_\shortparallel})$.  
On the contrary, the exact functional  will generally involve terms where, for example, 
there are two or more integrals over ${\mathbf x_\shortparallel}$, and kernels depending of those variables, and products of  $\psi$ with different arguments.
However,  regardless of the non locality of the exact expression, it must become local  when the surfaces
are sufficiently smooth and  close to each other.  Indeed, if the PFA becomes valid asymptotically in that limit, 
then the energy must approach a result which is a local function of $\psi$. Not whatever local functional but just one 
without derivatives.  

The way we found to depart slightly but significantly from the PFA, has been to add terms  involving derivatives
of $\psi$.  Namely, we shall assume that the electrostatic energy can be expanded in local terms involving 
derivatives of $\psi$.  One can think of the condition $\vert\nabla\psi\vert\ll 1$, as introducing a small, 
dimensionless expansion parameter.  In physical terms, this means that  the curved surface is 
almost parallel  to the plane on the points where it is satisfied. 

To introduce the first departure  from the PFA, we include terms with up to two derivatives. Then the 
electrostatic energy has to be  (up to this order) of the form:
\begin{equation} U_{\rm DE}\simeq\int
	d^2{\mathbf x_\shortparallel} \left[V(\psi)+Z(\psi)\vert\nabla\psi\vert^2\right]\,
	, \end{equation} 
for some functions $V$ and $Z$. The gradient is the two-dimensional one, and it can only appear in such a way that the energy is a scalar ($\psi$ is a scalar under changes of coordinates on the plane).
Besides, recalling  the equations of electrostatics, and on dimensional grounds, the result must 
be proportional to $\epsilon_0 V^2$. On top of that it  must reproduce $U_{\rm PFA}$ for constant $\psi$. Furthermore, as $\psi$ is the only other dimensionful quantity,  both functions $V$ and $Z$ have to be proportional to $\psi^{-1}$. Thus, we have restricted even further the functional to:
\begin{equation}
U_{\rm DE}\simeq \frac{\epsilon_0V_0^2}{2}\int d^2{\mathbf x_\shortparallel}\,  \frac{1}{\psi}(1+\beta_{\rm E}\vert\nabla\psi\vert^2)\, ,
\label{PFAimproved}
\end{equation}
where $\beta_{\rm E}$ is a numerical coefficient to be determined (the subindex $\rm E$ stands for electrostatics). It is worth stressing  that it is independent of the specific surface being considered,  as long 
as it is smooth. Therefore, it can be obtained once and for all just from its evaluation 
for a particular case. A simple procedure to obtain  the coefficient $\beta_{\rm E}$, when an exact analytic solution to the problem is known, would be to retrieve its value by  expanding that solution.  Let us do that for the configuration of a cylinder in front of a plane.  Inserting Equation (\ref{psicyl}) into Equation (\ref{PFAimproved}), and  performing the  integrals,  an expansion of the result in powers of $a/R$, allow us to fix $\beta_{\rm E}$.  Indeed, in order to agree with the expansion of the exact result in Equation (\ref{cpntlo}),  this fixes its value to $\beta_{\rm E}=1/3$.  
Of course, one will obtain the same value for any other particular example for which the exact solution  was known.

It is worthy of noting that, since the DE is a perturbative approach, it should be desirable to have a perturbative method to calculate the coefficient $\beta_{\rm EM}$. In other words, to compute it from first principles, using the appropriate expansion. One can do that, for instance, by solving perturbatively the Laplace equation  and then resorting to the method described  in Section \ref{sec:cas}.  We have performed that calculation in Ref.\cite{FLMelectro}, and  refer the reader to that work for details,
and also for the application of the DE to other  electrostatic examples. 

\subsection{Two Smooth Surfaces}\label{2_surf_elect}

 As a natural generalization of the previously discussed situation, let us now  consider two surfaces
described by the two functions $\psi_1({\mathbf x_\shortparallel})$ and  $\psi_2({\mathbf x_\shortparallel})$, 
each one of them  measuring the respective  height of a  surface with respect to a reference plane 
$\Sigma$. This geometry was first considered in the context of the DE for the Casimir effect in Ref.\cite{Bimonte1}.

To construct the DE for the electrostatic energy  in this case,  we keep up to two derivatives of the functions. This  allows we  to write the general expression:
\begin{eqnarray}\label{2surfaces}
U_{\rm DE}[\psi_1,\psi_2] &=&\int_{\Sigma} d^2{\mathbf x_\shortparallel}\, 
U_\shortparallel({\psi)}\, \left[1+\beta_1 
\vert\nabla \psi_1\vert^2 \right.  \nonumber \\
&+&\beta_2\vert \nabla \psi_2\vert^2 +
\beta_\times\nabla \psi_1\cdot\nabla \psi_2 \nonumber \\
&+& \left.\beta_{\rm -} \, {\hat {\mathbf z}} \cdot
\nabla \psi_1  \times \nabla \psi_2) + \cdots\right],
\end{eqnarray}
where $\psi=\vert \psi_2-\psi_1\vert$ is the height difference, $U_\shortparallel(\psi)=\epsilon_0V_0^2/(2\psi)$ is the electrostatic energy
between parallel plates, and the dots denote higher derivative terms.  Equation \eqref{2surfaces} actually contains four 
numerical constants: $\beta_1$, $\beta_2$, $\beta_\times$, and $\beta_-$.  However, symmetry considerations imply 
some constraints on them: the energy must be invariant under the interchange of $\psi_1$ and
$\psi_2$, since that is just a relabeling: $\beta_1=\beta_2$ and $\beta_-=0$. Furthermore, in order to
reproduce the  result for a single smooth surface in front of a plane we must have  $\beta_1=\beta_2=1/3$. The coefficient $\beta_\times$ can be determined taking into account that the
energy should be invariant 
under a simultaneous rotation of both
surfaces \cite{Bimonte1}. Indeed, for an infinitesimal rotation of each surface by an angle
$\epsilon$ in the plane $(x,z)$, the changes induced on the functions
$\psi_i$ 
are
$\delta\psi_i=\epsilon(x+\psi_i\partial_x\psi_i)$, for $i=1,2$. To
simplify the determination of $\beta_\times$ we can assume that, initially, $\psi_1=0$ and that
$\psi_2$ is only a function of $x$. Computing explicitly the variation of
$U_{\rm DE}$ to linear order in $\epsilon$ one can show that
\begin{equation}
\delta U_{\rm DE}=0\Rightarrow \beta_\times={1}/{3}\, , 
\end{equation}
and therefore

\begin{widetext}
    
\begin{equation}\label{2surfacesfinal}
U_{\rm DE}[\psi_1,\psi_2]=\frac{\epsilon_0V_0^2}{2}\int_{\Sigma} d^2{\mathbf x_\shortparallel}\,\frac{1}{\psi}\, \left[1+\frac{1}{3}\left(
\vert\nabla \psi_1\vert^2 +\vert \nabla \psi_2\vert^2 +  \nabla \psi_1\cdot\nabla \psi_2\right)
\right].
\end{equation}
\end{widetext}

Note that, by taking the variation of the electrostatic energy Equation (\ref{2surfaces})
with respect to translations or rotations of one of the surfaces, one can obtain  the vertical  and lateral components of the force, as well as the torque, due to the remaining surface.  

The identities $\beta_1=\beta_2$ and $\beta_-=0$  are universally valid, regardless of the interaction (as long as the surfaces are of an identical nature), but $\beta_\times =\beta_1$ holds true for the electrostatic interaction. This depends upon the fact that the leading term is proportional to  $\psi^{-1}$  (i.e. it is then not valid for the Casimir energy). 

For later use, let us recall that, for a general function $E_\shortparallel(\psi)$, the relation between the different coefficients becomes \cite{Bimonte1}:
\begin{equation}\label{betax}
2(\beta_1+\beta_2)+ 2\beta_\times+\psi\frac{d\log E_\shortparallel}{d\psi}=1\, .
\end{equation}
The relation {\eqref{betax}} shows that, for any interaction, the DE for the interaction energy between two curved surfaces can be reduced to the problem of a single surface in front of a plane. Indeed, in the later case one can determine $\beta_1$ and $\beta_2$, while Equation \eqref{betax} determines the remaining coefficient $\beta_\times$.

To summarize: when computing the electrostatic  energy associated with a configuration of two conductors at different potentials, with smoothly curved surfaces,  one can go beyond the PFA by simply assuming that the energy admits an expansion in  derivatives of the functions that define the shapes of the conductors.
If the  exact electrostatic energy for a single non trivial curved configuration is known, one can determine all the  free parameters in the expansion. 

Finally, the NTLO correction produces an appreciable  improvement in the DA and, by the same token, also provides an assessment for its validity. 
An interesting alternative approach  to compute electrostatic forces beyond the PFA can be found in Ref.~\cite{Hudlet}.
\section{Obtaining the DE from a Perturbative Expansion}\label{sec:sum}

Regardless of the interaction considered, the DA and its improvement, the DE, can  be obtained by performing the proper resummation of  a perturbative expansion \cite{FLMrev2014}. The required  expansion is in powers of the departure of the surfaces, about a two flat parallel planes configuration.
This connection yields a systematic and quite general approach to obtain the DE, even when an exact solution
is not available.  

To keep things general, we work  with a general functional of the surface; that functional may
correspond to an energy, free energy, force, etc.  Besides, we do not make  any assumption about 
the kind of interaction involved,  not even about whether it satisfies a superposition principle or not.

To begin, let us we assume a geometry where there are two surfaces,  one of which, $L$, is a plane, which with a proper choice of Cartesian coordinates ($x_1$, $x_2$, $x_3$), is described by $x_3 = 0$.  The other one, $R$,  is assumed to be describable by $x_3 = \psi({\mathbf x_\shortparallel})$.

The object for which we  implement the approximation is denoted by $F[\psi]$, a functional of  $\psi$.  Then we note 
that the PFA for $F$, to be denoted here by $F_0$, is obtained as follows: add, for each ${\mathbf x_\shortparallel}$, the product of a local surface density ${\mathcal F}_0(\psi({\mathbf
x_\shortparallel}))$ depending only on the value of $\psi$ at the point ${\mathbf x_\shortparallel}$, times the surface element area; namely,
\begin{equation}
F_0 [\psi]\;=\; \int d^2{\mathbf x_\shortparallel}
\, {\mathcal F}_0(\psi({\mathbf x_\shortparallel})) \;. 
\end{equation} 
The surface density is, in turn, determined by the (assumed) knowledge  of the exact form 
of $F$ for the case of two parallel surfaces, as follows:
\begin{equation}
{\mathcal F}_0(a) \;=\; \lim_{{\mathcal S} \to \infty} 
\big[\frac{F[a]}{\mathcal S} \big]\;, 
\end{equation} 
where ${\mathcal S}$ denotes the area of the $L$ plate and $a$ is a constant. Namely, to determine the density one 
needs to know the functional $F$ just for  constant functions $\psi \equiv a$. Note that, if the functional $F$ is the interaction energy between  the surfaces, ${\mathcal F}_0$ becomes the interaction energy per unit area $E_{\shortparallel}$, and $F_0$ becomes  $U_{\rm PFA}$
(see Equation (\ref{PFA})).

Let us now show how to derive the PFA (and its corrections) by the resummation of a perturbative expansion. To that end, we evaluate $F$ for a $\psi$ having the form: 
\begin{equation}
\psi({\mathbf x_\shortparallel}) = a + \eta( {\mathbf
x_\shortparallel})\; .
\end{equation} 
and write the resulting perturbative expansion in powers of $\eta$, which has the general form:

\begin{widetext}
\begin{equation}
F[\psi]={\mathcal S\mathcal F_0}(a)+\sum_{n\geq 1}\int \frac{d^2k_\shortparallel^{(1)}}{(2\pi)^2}
...\frac{d^2k_\shortparallel^{(n)}}{(2\pi)^2} \,
\delta(k_\shortparallel^{(1)}+...+k_\shortparallel^{(n)}) \, h^{(n)}(k_\shortparallel^{(1)},...,k_\shortparallel^{(n)})
\, {\widetilde{\eta}}(k^{(1)}_\shortparallel) ...{\widetilde{\eta}}(k^{(n)}_\shortparallel)\; ,
\label{genexp}
\end{equation}
\end{widetext}
where $\delta(\cdot )$ is the Dirac delta function,  and the form factors $h^{(n)}$ can be computed by using  perturbative techniques. For the Dirichlet-Casimir
effect, this can be done  in a rather systematic way~\cite{4thorder}. Although the  approach to follow in order 
to obtain those form factors may  depend strongly on the kind of system considered, the form of the expansion shall be the same.  Note that the form factors may depend on $a$, although,  in order to simplify the notation,  we will not make that dependence explicit.

Up to now, we have not used  the hypothesis of smoothness of the $R$ surface.
We do that now by assuming that the Fourier transform $\widetilde\eta$ is peaked at the zero momentum. What follows is to make use of this assumption for all terms in the expansion.
In Equation (\ref{genexp}), we set then: $h^{(n)}(k_\shortparallel^{(1)},...,k_\shortparallel^{(n)})\simeq h^{(n)}(0,...,0)$, and, as a consequence:
\begin{equation}
F(\psi)\simeq {\mathcal S \mathcal F_0}(a)+\sum_{n\geq
1}h^{(n)}(0,..,0)\int d^2{\mathbf x_\shortparallel} [\eta({\mathbf
x_\shortparallel})]^n \;.
\label{low}
\end{equation}
One could evaluate the form factors at the zero momentum straighforwardly. However, there is a shortcut here that allows one to obtain all of them immediately: consider  a constant  $\eta({\mathbf x_\shortparallel})=\eta_0$, so that the interaction energy is given 
by Equation (\ref{low}) with the replacement $\int
d^2{\mathbf x_\shortparallel} \,\eta({\mathbf x_\shortparallel})^n\rightarrow
{\mathcal S}\eta_0^n$. For this particular case, $F$ becomes  just 
the functional corresponding to parallel plates, which are separated by a distance $a+\eta_0$:
\begin{equation}
{\mathcal F_0}(a+\eta_0)= {\mathcal F_0}(a) + \sum_{n\geq 1}h^{(n)}(0,..,0) \eta_0^n\, .
\label{F_0fin}
\end{equation}
We then conclude that, in this low-momentum approximation, the series can be
summed up with the result: 
\begin{equation}
F_0[\psi]\simeq  
\int d^2{\mathbf x_\shortparallel} {\mathcal F_0}\left (a+\eta({\mathbf x_\shortparallel})\right)= 
\int d^2{\mathbf x_\shortparallel} {\mathcal F_0}(\psi)\, ,
\end{equation} 
which is just the PFA.

The calculation just above  shows that, for the class of geometries considered in this paper, the PFA 
can be justified  from first principles as the result of  a resummation of a perturbative calculation 
corresponding to almost flat surfaces. 
In order to be well defined, 
the PFA requires that the form factors
$h^{(n)}(k_\shortparallel^{(1)},...,k_\shortparallel^{(n)})$ have a finite limit as $k_\shortparallel^{(i)}\to 0$.

This procedure also suggests how  the PFA could be improved; one can include
the NTLO terms in the low-momentum expansions of the form factors.  We assume that they can be expanded
in powers of the momenta up to the second order. We stress that this is by no means a  trivial assumption. Indeed, 
depending on the the interaction considered, the form factors could include 
nonanalyticities (we will discuss some explicit examples below). 
In case of no nonanalyticities, one can introduce the expansions:
\begin{widetext}
\begin{equation} \label{eq:hexp}
 h^{(n)}(k_\shortparallel^{(1)},...,k_\shortparallel^{(n)})= h^{(n)}(0,...,0) +
\sum_{i,\alpha}A^{(n)}_{i\alpha}k_{\shortparallel\, \alpha}^{(i)}+
\sum_{i,j,\alpha,\beta}B^{(n)}_{ij\alpha\beta}
 k_{\shortparallel\, \alpha}^{(i)}  k_{\shortparallel\, \beta}^{(j)}\,\ldots , 
\end{equation}
\end{widetext}
 for some $a-$dependent coefficients $A^{(n)}_{i\alpha}$ and
$B^{(n)}_{ij\alpha\beta}$. Here $i,j=1,...,n$ label arguments while $\alpha,\beta = 1,2 $ label their components. Symmetry considerations are crucial, since they allow us to simplify the above expression  \eqref{eq:hexp}, as follows: rotational invariance implies that the form factors depend only on the scalar products $k_\shortparallel^{(i)}\cdot k_\shortparallel^{(j)}$. Additionally, they have to  be symmetric under the interchange of any two momenta. This thus leads to 
\begin{widetext}
\begin{equation}
 h^{(n)}(k_\shortparallel^{(1)},...,k_\shortparallel^{(n)})= h^{(n)}(0,...,0) + B^{(n)}\sum_i k_\shortparallel^{(i)\, 2}+ C^{(n)}\sum_{i\neq j}k_\shortparallel^{(i)}\cdot k_\shortparallel^{(j)} \, ,
 \label{cuad}
  \end{equation}
\end{widetext}
for some coefficients $B^{(n)}$ and $C^{(n)}$. 

Inserting Equation \eqref{cuad} into Equation \eqref{genexp} and taking integrations by parts, one then
finds the form of the first correction to the PFA:

\begin{equation}
F_2[\psi]=\int d^2{\mathbf x_\shortparallel}\, \left[\sum_{n\geq 2}D^{(n)}\,
\eta^{n-2}\right] |\nabla \eta |^2\;,
\label{orden2}
\end{equation}
where the coefficients $D^{(n)}$ are linear combinations of $B^{(n)}$ and $C^{(n)}$. The subindex  $2$ in $F$ indicates that this is the part of the functional containing two derivatives.

We complete the calculation by calculating the sum in Equation \eqref{orden2}. To that end, we evaluate the correction 
$F_2$ for  a particular case: $\eta({\mathbf x_\shortparallel})= \eta_0 + \epsilon({\mathbf x_\shortparallel})$, with $\epsilon\ll\eta_0$, and expand up to the second order in $\epsilon$. Thus,
\begin{equation}
F_2[a+\eta_0+\epsilon]=\int d^2{\mathbf x_\shortparallel}\, \left[\sum_{n\geq
2}D^{(n)}\, \eta_0^{n-2}\right] |\nabla \epsilon |^2\, .
\label{orden2eps}
\end{equation}
The resummation can be obtained in this case, by considering the usual perturbative evaluation of the interaction
energy up to second order in $\epsilon$. This evaluation does, naturally, depend on the interaction considered, but, once one
has that result  one can obtain the sum  of the series above. We  we will denote by $Z$ that sum, namely:
\begin{equation}
Z(a+\eta_0)\equiv \sum_{n\geq 2}D^{(n)}\, \eta_0^{n-2}\, .
\label{zeta2}
\end{equation}
Upon replacement $\eta_0\to\eta$ in Equation (\ref{zeta2}), one obtains
\begin{equation}
F_2[\psi]=\int d^2{\mathbf x_\shortparallel}\, Z(\psi) |\nabla \psi |^2\, .
\label{fin}
\end{equation}
This is the NTLO correction to the PFA. 
This concludes our systematic derivation of the PFA, including its first correction, a result which may be put as follows:
\begin{equation}\label{eq:de}
F_{\rm DE}[\psi] \;=\; \int d^2{\mathbf x_\shortparallel} \, \Big[ V(\psi) \, + \,
 Z(\psi) |\nabla \psi |^2 \Big] \;, 
\end{equation}
where $V(\psi)={\mathcal F}_0(\psi)$ is determined from the  (known) expression for the interaction energy between parallel surfaces, while
$Z(\psi)$ can be computed using a perturbative technique. In practice,   $Z(\psi)$ can be evaluated  setting
$\eta_0=0$ in Equation (\ref{zeta2}).

The higher orders may be derived by an extension of the procedure described just above. It
should be evident that,  for the expansion to be well-defined,  the
analytic structure of the form factors  is quite relevant. Indeed, the existence of nonanalytic
zero-momentum contributions  can  render the DE non applicable. This should be expected on physical grounds, since the presence of
nonanalytic terms implies that the functional cannot be approximated, in
coordinate space, by the single integral of a local density. Physically, it
is a signal that the nonlocal aspects of the  interaction cannot be ignored.  That should not come up
as a surprise, when one recalls that the same kind of phenomenon does happen
when evaluating the effective action in quantum field theory, and the quantum effects
contain contributions due to virtual  massless particles. In this case, the effective 
action may develop nonanalyticities at zero momentum.

The main messages of this Section are the following: irrespective of the nature of the interaction, the energy and forces  between objects are  functionals of their shapes. The PFA is recovered when the form factors of the functionals are evaluated at zero momentum. 
Enhancements to this approximation are achievable by expanding these form factors at low momenta.
 If the expansion is analytic, a resummation of the form factors produces the DE.

\section{DE for the Zero-Temperature Casimir Effect}\label{sec:cas}
 
 The application of the DE to the Casimir interaction energy between two objects
 was, actually, our original motivation to introduce the approximation, and 
 it is useful briefly review some aspects of this application here.
 We consider first a real vacuum 
 scalar field satisfying Dirichlet boundary conditions {(Section \ref{sec51})}
and then we move to the  EM  
 field with perfect-conductor boundary conditions {(Section \ref{sec52})}. 
We follow Ref.~\cite{FLM2011} for the
derivation of the DE in the Dirichlet case.

\subsection{Scalar Field with Dirichlet Boundary Conditions}\label{sec51}
We  consider here  a massless real scalar field $\varphi$ in $3+1$ dimensions, coupled to two mirrors which impose
Dirichlet boundary conditions.  In our Euclidean conventions, we use $x_0,x_1,x_2,x_3$ to denote the 
spacetime coordinates, $x_0$ being the imaginary time.
As before, the mirrors occupy two surfaces, denoted by $L$ and $R$, defined by the equations $ x_3 = 0 $ and  $x_3 =  \psi({\mathbf x_\shortparallel})$,  respectively. 

On only dimensional  grounds, and using natural units ($\hbar \equiv c  \equiv 1$), 
the DE approximation to the interaction energy to be of the form
\begin{equation}
E_{\rm DE}= -\frac{\pi^2}{1440} \int d^2{\mathbf x_\shortparallel} \;
\frac{1}{\psi^3}\left[\alpha_{\rm D}+\beta_{\rm D}(\partial_\alpha\psi)^2\right],
\label{EDEdimen}
\end{equation}
where $\alpha_{\rm D}$ and $\beta_{\rm D}$ are dimensionless coefficients that do not depend on the geometry. 
The subindex $D$ stands for Dirichlet. An 
evaluation of the above expression for parallel plates fixes $\alpha_D\equiv 1$.  As in the electrostatic case, the coefficient $\beta_{\rm D}$ could be computed from explicit examples where the interaction energy is known exactly. 

Let us recall, from Section \ref{sec:sum}, that the interaction energy can also be computed from an expansion of the Casimir energy in powers of $\eta$ for
\begin{equation}
	\psi({\mathbf x_\shortparallel}) \;=\; a \,+\, \eta({\mathbf
	x_\shortparallel}) \;,
\label{psi_pert}
\end{equation}
where $a$ (assumed to be greater than zero) is the spatial average of $\psi$ whereas $\eta$ contains 
its varying piece. The expansion needed is of the second order in $\eta$, and with up to two spatial 
derivatives.

To obtain such an expansion, we start from a rather general yet formal  expression for the energy 
(for earlier perturbative computations of the Casimir force see, for example, Ref.\cite{earlier-pert1,earlier-pert2}). 
That formal expression follows from the functional approach to the Casimir effect, where we  
deal with   ${\mathcal Z}$, the  zero-temperature limit of a partition function.  That partition function, for a scalar field 
in the presence of  two Dirichlet mirrors is given by
\begin{equation}
	{\mathcal Z}\;=\; \int {\mathcal D}\varphi \;\delta_L(\varphi) \,
	\delta_R(\varphi) \; e^{- S_0(\varphi) } \;,
\end{equation}
with $S$ denoting the real scalar field free (Euclidean) action
\begin{equation}
S_0(\varphi) \;=\; \frac{1}{2} \, \int d^4 x \, (\partial \varphi)^2
\;,
\end{equation}
while the $\delta_L$ and $\delta_R$  impose Dirichlet boundary conditions on the $L$ and $R$ surface, respectively.

The vacuum energy, $E$, is then obtained as follows:
\begin{equation}\label{eq:evac}
	E \;=\;- \, \lim_{T \to \infty} \log {\mathcal Z}/T \;,
\end{equation}
where $T$ is the extent of the time dimension (or $\beta^{-1}$, in a thermal partition function setting).  We 
discard from $E$ the terms that do not contribute  to the Casimir interaction energy between 
the two surfaces. These terms will appear as factors in ${\mathcal Z}$; among them the one describing the zero point  energy  of 
the field in the absence of the plates, and  also the `self-energy' contributions, due to the vacuum 
distortion produced by each mirror, even when the other is infinitely far apart. 

Exponentiating the two Dirac delta functions by introducing two auxiliary fields,
$\lambda_L$ and $\lambda_R$, we obtain for ${\mathcal Z}$ an equivalent expression:
\begin{equation}
{\mathcal Z}\;=\; \int {\mathcal D}\varphi {\mathcal D}\lambda_L
\, {\mathcal D}\lambda_R \; e^{-S(\varphi;\lambda_L,\lambda_R)} \;,
\end{equation}
with
\begin{widetext}
\begin{equation}
S(\varphi;\lambda_L,\lambda_R) =S_0(\varphi) -i 
\int d^4x \varphi(x) 
\left[ 
\lambda_L(x_\shortparallel) \delta(x_3) 
+ \lambda_R(x_\shortparallel) \sqrt{g_R({\mathbf x_\shortparallel})} \, 
\delta(x_3-\psi({\mathbf x_\shortparallel})) \right]
\nonumber \end{equation}
\end{widetext}
where we have introduced $x_\shortparallel\equiv (x_0,x_1,x_2) = (x_0, {\mathbf x_\shortparallel})$. The factor depending on the determinant of the 
induced metric on the $R$, $g_R ({\mathbf x_\shortparallel}) \equiv 1 + |\nabla\psi({\mathbf x_\shortparallel})|^2$  makes the expression above reparametrization invariant. However, by a redefinition of the auxiliary field $\lambda_R$ one gets rid of that factor, at the expense of generating a 
Jacobian. That Jacobian does not depend on the distance between the two surfaces, since only derivatives
of $\psi$ are involved.  Therefore it will not contribute the the Casimir interaction energy and thus we 
shall subsequently ignore such factor, as well as others that will appear in the course of the calculations.

Integrating out $\varphi$, we see that ${\mathcal Z}_0$, corresponding to the field $\varphi$ in the absence of boundary conditions
factors out, while the rest becomes an integral over the auxiliary fields:
\begin{equation}
{\mathcal Z}={\mathcal Z}_0  
\int {\mathcal D}\lambda_L  {\mathcal D}\lambda_R  
e^{-\frac{1}{2} \int d^3x_\shortparallel \int d^3y_\shortparallel 
\sum_{\alpha,\beta}\lambda_\alpha(x_\shortparallel)  
{\mathbb T}_{\alpha\beta} 
\lambda_\beta(y_\shortparallel)},
\end{equation}
with:
\begin{equation}
{\mathcal Z}_0\;=\; \int {\mathcal D}\varphi \; e^{-S_0(\varphi)} \;,
\end{equation}
and $\alpha,\,\beta=L,R$ . We have introduced the objects:
\begin{align}
{\mathbb T}_{LL}(x_\shortparallel,y_\shortparallel) &=\langle
x_\shortparallel,0|(-\partial^2)^{-1} |y_\shortparallel,0\rangle \\
{\mathbb T}_{LR}(x_\shortparallel,y_\shortparallel) &=\langle
x_\shortparallel,0|(-\partial^2)^{-1}
|y_\shortparallel,\psi({\mathbf y_\shortparallel})\rangle \\
{\mathbb T}_{RL}(x_\shortparallel,y_\shortparallel) &=\langle
x_\shortparallel,\psi({\mathbf x_\shortparallel})|(-\partial^2)^{-1} 
|y_\shortparallel,0\rangle \\
{\mathbb T}_{RR}(x_\shortparallel,y_\shortparallel) &=\langle
x_\shortparallel,\psi({\mathbf x_\shortparallel})|(-\partial^2)^{-1} 
|y_\shortparallel,\psi({\mathbf y_\shortparallel})\rangle 
\end{align}
where we use a ``bra-ket'' notation to denote matrix elements of operators,
and $\partial^2$ is the four-dimensional Laplacian. Thus, for example,
\begin{equation}
	\langle x|(-\partial^2)^{-1} |y\rangle \,=\, \int
	\frac{d^4k}{(2\pi)^4} \, \frac{e^{i k \cdot (x-y)}}{k^2}\;.
\end{equation}
A subtraction of the zero point contribution contained in ${\mathcal Z}_0$ leads to:
\begin{equation}
E \;=\; \lim_{T \to \infty} \big(
\frac{1}{2 T} {\rm Tr}  \log {\mathbb T} \big)\;,
\end{equation}
which still contains self-energies. Up to now, we have obtained a formal expression for
the vacuum energy; let us now proceed to evaluate its DE.

We need to expand  $E$ to the second order in $\eta$, keeping up to the second order term in an 
expansion in derivatives. It is convenient to do so first for  $\Gamma\;\equiv\; \frac{1}{2} {\rm Tr}  \log {\mathbb T}$ . Namely,
\begin{equation}\label{eq:expansion1}
	\Gamma(a,\eta)\;=\; \Gamma^{(0)}(a) \;+\;\Gamma^{(1)}(a,\eta) \;+\;
 \Gamma^{(2)}(a,\eta) \;+\;\ldots
\end{equation}
where the upper index denotes the order in derivatives. Each term will be a certain coefficient times 
the spatial integral over ${\mathbf x_\shortparallel}$ of a local term, depending on $a$ and on $\eta$-derivatives. Additionally, because
the configuration is time-independent, they should be proportional to $T$ (a factor that will 
cancel out).
Expanding first the matrix ${\mathbb T}$ in powers of $\eta$
\begin{equation}
{\mathbb T}={\mathbb T}^{(0)}+{\mathbb T}^{(1)}+{\mathbb T}^{(2)}+\ldots\, ,
\end{equation}
we obtain: $\Gamma \,=\,
\Gamma^{(0)} + \Gamma^{(1)}+ \Gamma^{(2)}\,+\,\ldots$,
\begin{eqnarray}
\Gamma^{(0)}&=&\frac{1}{2} {\rm Tr}  \log {\mathbb T}^{(0)}\nonumber\\
\Gamma^{(1)}&=&\frac{1}{2} {\rm Tr}  \log \left[({\mathbb T}^{(0)})^{-1}
{\mathbb T^{(1)}}\right]\nonumber\\
\Gamma^{(2)}&=&\frac{1}{2} {\rm Tr}  \log \left[({\mathbb T}^{(0)})^{-1}{\mathbb T}^{(2)}\right] \nonumber \\
&-& \frac{1}{4} {\rm Tr}  \log \left[ ({\mathbb T}^{(0)})^{-1}{\mathbb T}^{(1)}  
({\mathbb T}^{(0)})^{-1}{\mathbb T}^{(1)}
\right]\;,
\end{eqnarray}
where, in $\Gamma^{(l)}$, we need to keep up to $l$ derivatives of $\eta$.

Then, the  zeroth-order term is obtained as follows: replace $\psi$ by a constant, $a$, and then subtract from 
the result its  $a
\to \infty$ limit (this gets rid of self-energies). This leads to:
\begin{equation}
	\Gamma^{(0)}(a) \;=\; \frac{1}{2} \, {\rm Tr} \log \big[ 1 - 
	(T_{LL}^{(0)})^{-1} T_{LR}^{(0)}
	(T_{RR}^{(0)})^{-1} T_{RL}^{(0)} \big] \;.
\end{equation}
Here, the $T_{\alpha\beta}^{(0)}$ are identical to the ones for two flat parallel mirrors separated by  a distance $a$ .   

Taking the trace, leads to:
\begin{equation}
\Gamma^{(0)}\,=\, \frac{T}{2} \, \int d^2{\mathbf x_\shortparallel}\int\frac{d^3k_\shortparallel}{(2\pi)^3}
\log[1-e^{-2k_\shortparallel a}]\, .
\end{equation}
Then, we recall the general derivation to note that the replacement $a \to \psi$  leads to: 
\begin{eqnarray}
E^{(0)} &=& \frac{1}{2} \, \int d^2{\mathbf x_\shortparallel}\int\frac{d^3k_\shortparallel}{(2\pi)^3}
\log[1-e^{-2k_\shortparallel\psi({\mathbf x_\shortparallel})}] \nonumber \\
&=& -\frac{\pi^2}{1440} \int d^2{\mathbf x_\shortparallel}\; \frac{1}{\psi({\mathbf x_\shortparallel})^3}  ,
\end{eqnarray}
which is the PFA expression for the vacuum energy.

To improve on the previous result, we consider its first non trivial correction.  There can be no first order term because of symmetry considerations. while to terms contribute to the second order
\begin{equation}
	\Gamma^{(2)} \;=\; \Gamma^{(2,1)} \,+\, \Gamma^{(2,2)} 
\end{equation}
where,
\begin{equation}
\Gamma^{(2,1)} \,=\, \frac{1}{2} {\rm Tr}  \log \left[({\mathbb
T}^{(0)})^{-1}{\mathbb T}^{(2)}\right] 
\end{equation}
and
\begin{equation}
\Gamma^{(2,2)} \,=\,-\frac{1}{4} {\rm Tr}  \log \left[ ({\mathbb
T}^{(0)})^{-1}{\mathbb T}^{(1)}  ({\mathbb T}^{(0)})^{-1}{\mathbb T}^{(1)}
\right] \;.
\end{equation}

In the terms above, we have to keep just up to two derivatives of $\eta$. We see that, in Fourier space, and before implementing any expansion in momentum (derivatives),  they have the structure:
\begin{equation}
	\Gamma^{(2,j)}\;=\; \frac{T}{2} \, \int \frac{d^2{\mathbf k_\shortparallel}}{(2\pi)^2}
	f^{(2,j)}({\mathbf k_\shortparallel}) \, |\tilde{\eta}({\mathbf k_\shortparallel})|^2 
\end{equation}
($j=1,2$), with  $\tilde{\eta}$  denoting the Fourier transform of $\eta$, and with  the $f^{(2,j)}$ kernels denoting 
the  $k_0\to 0$ (i.e., static) limits  of the more general expressions:
\begin{eqnarray}
f^{(2,1)}(k_\shortparallel) &=& - \int \frac{d^3p_\shortparallel}{(2\pi)^3} \frac{|p_\shortparallel|
|p_\shortparallel+k_\shortparallel|}{1	- e^{- 2 |p_\shortparallel + k_\shortparallel| a}} 
\nonumber\\
f^{(2,2)}(k_\shortparallel) &=& - \int \frac{d^3p_\shortparallel}{(2\pi)^3} 
\frac{|p_\shortparallel| |p_\shortparallel+k_\shortparallel| e^{-2 |p_\shortparallel+k_\shortparallel| a} 
( 1 + e^{-2 |p_\shortparallel| a}) }{(1 - e^{- 2 |p_\shortparallel| a}) 
(1 - e^{- 2 |p_\shortparallel + k_\shortparallel| a})}
\nonumber .
\end{eqnarray}
By subtracting  all the $a$-independent contributions,  one finds:
\begin{equation}
	\Gamma^{(2)}\;=\; \frac{T}{2} \,\int \frac{d^2{\mathbf k_\shortparallel}}{(2\pi)^2}
	f^{(2)}({\mathbf k_\shortparallel}) \, |\tilde{\eta}({\mathbf k_\shortparallel})|^2 
\end{equation}
with:
\begin{equation}
f^{(2)}(k_\shortparallel) \;=\; - 2 \int \frac{d^3p_\shortparallel}{(2\pi)^3} 
\frac{|p_\shortparallel| \,|p_\shortparallel+k_\shortparallel|}{(1 - e^{- 2 |p_\shortparallel| a}) (e^{2 |p_\shortparallel+k_\shortparallel| 
a} - 1)}\, .
\end{equation}

The low-momentum behaviour of $f^{(2)}$ determines whether the DE can be applied or not. 
In this case, the function is analytic and therefore a local expansion of the vacuum energy exists. 
We need to extract its ${\mathbf k}^2$ order term in a  Taylor expansion at zero momentum, namely $f^{(2)}({\mathbf k_\shortparallel}) \;\simeq\; \chi  \, {\mathbf k_\shortparallel}^2$. We find:
\begin{equation}
	\chi \,=\,  - \frac{\pi^2}{1080 \, a^3} \;.
\end{equation}
Thus,
\begin{eqnarray}
\Gamma^{(2)}(a,\eta) &=&- \frac{T}{2} 
\frac{\pi^2}{1080} 
\int \frac{d^2{\mathbf k_\shortparallel}}{(2\pi)^2}
\frac{{\mathbf k_\shortparallel}^2}{a^3} \, |\tilde{\eta}({\mathbf k_\shortparallel})|^2 \nonumber\\
&=&-\frac{T}{2} \frac{\pi^2}{1080} \int d^2{\mathbf x_\shortparallel}\,
\frac{1}{a^3} \, (\partial_\alpha \eta)^2 \;.
\end{eqnarray}
Therefore, the NTLO term in the DE becomes:
\begin{equation}
E^{(2)}=\frac{\Gamma^{(2)}(\psi)}{T} \,= \, -\frac{1}{2} 
\frac{\pi^2}{1080} \int d^2{\mathbf x_\shortparallel}\,
\frac{(\partial_\alpha \psi)^2}{\psi^3} \;,
\end{equation}
where the index $\alpha$ runs from $1$ to $2$.

Putting together the terms up to second order,
\begin{equation}
E_{\rm DE} \equiv 
E^{(0)}+E^{(2)}=- \frac{\pi^2}{1440}\int d^2{\mathbf x_ \shortparallel} \;
\frac{1}{\psi^3}\left[1+\frac{2}{3}(\partial_\alpha\psi)^2\right].
\label{final}\end{equation}
The leading-order term above is the Casimir energy according to the PFA , while the second order one 
represents the first significant deviation from it. We note that the structure of both terms had been 
anticipated by dimensional analysis and symmetry considerations. The overall normalization, on the 
other hand, had been fixed by our  previous knowledge of the (well-established) result for parallel plates. 

We would like to insist on the fact that the relative weight between the PFA and its correction term--the factor $\beta_{\rm D} = 2/3$--is independent of the surface geometry. This value of $\beta_{\rm D}$ has been independently
corroborated in concrete examples by expanding the exact Casimir energy expressions. Interesting 
cases among them are, for example, either a sphere or a cylinder positioned in front of a plane.

We conclude this Section with an application of the DE to the particular geometry 
of a sphere in front of a plane.  Let us express the function $\psi$ of Equation (\ref{psicyl}) in polar 
coordinates $\rho$, with  $R$  the radius of the sphere and $d$ the distance to the plane. 
The function $\psi(\rho)$ describes an hemisphere when $0\leq \rho \leq R$. By inserting the expression of 
$\psi$ into the DE for the Casimir energy, it becomes possible to explicitly calculate the integrals, to
get a rather compact analytical expression:
\begin{equation}
    E_{\rm DE} = E_{\rm PFA} \left( 1 + \frac{1}{3} \frac{a}{R}\right) \label{spD}, 
\end{equation}
where $E_{\rm PFA} = -\hbar c \pi^3 R/(1440 a^2)$.

\subsection{The EM Case}\label{sec52}
The results for the scalar field satisfying Dirichlet boundary conditions, described  in Section (\ref{sec51}) above,  have been
generalized to different boundary conditions and fields.  Results for the EM  field case and two curved 
surfaces have been presented in Ref.\cite{Bimonte1}. Note that, as pointed out  at the end of 
Section \ref{2_surf_elect},  symmetry considerations allow for the two-surface problem to be reduced to the 
one of a curved surface facing  a plane, namely, the geometry we have just  dealt with in the Dirichlet case 
above.  Indeed, as shown in  Ref. \cite{Bimonte1},  the extension of Ref.\cite{FLM2011} to two curved surfaces
is restricted among other things by the tilt invariance of the reference plane,  to which the two surfaces 
can be projected. This served as a rigorous test for the self-consistency of perturbative results. 

Venturing beyond the scalar  Dirichlet (D) case of Ref.\cite{FLM2011}, they calculated the DE for Neumann (N), 
mixed D/N, and electromagnetic (EM) (perfect metal) surfaces. Interestingly, they observed that the EM 
correction must align with the sum of D and N corrections. They also replicated previous findings for
cylinders under D, N, and mixed D/N conditions, as well as for the sphere with D boundary conditions. 
However, their calculations did not confirm previous results for the sphere/plane geometry, either with 
N or EM boundary conditions. Indeed, the results for $\beta$ were found to disagree with those obtained from 
Refs. \cite{bordag08,bordag101,bordag102}.  This discrepancy was later resolved in Ref.\cite{bordag-teo} in favour 
of the results in \cite{Bimonte1}.

Another interesting concrete example presented in \cite{Bimonte1} is the DE   for two spheres of radii
$R_1$ and $R_2$, both imposing the same boundary conditions.  It was found there that
\begin{equation}
    E = E_{\rm PFA} \left[1 - \frac{a}{R_1 + R_2} + (2 \beta -1) \left( \frac{a}{R_1} + \frac{a}{R_2}\right) \right], \label{bimonte}
\end{equation}
where $E_{\rm PFA} = -(\alpha \pi^3 R_1 R_2))/[1440a^2(R_1 + R_2)]$; $a$ is chosen to be the distance of closest separation, 
and $\beta$ is a number that depends on the type of boundary condition, as can be seen from Table \ref{betas}.
$\alpha = \alpha_{\rm EM}= 2$ in the EM boundary conditions case.  The corresponding formula for the sphere/plane case
can be obtained by taking one of the two radii to infinity (in fact it coincides with the D case in Equation (\ref{spD}) when $\alpha = \alpha_{\rm D} = 1$ and $\beta = \beta_{\rm D} = 2/3$).

A rather different example corresponds to two circular cylinders (with identical boundary conditions) 
whose axes are inclined at a relative angle $\theta$. Using the DE,  the interaction Casimir energy reads:
\begin{equation}\label{incl_cyl}
    E =-\frac{\alpha \pi^3\sqrt{R_1R_2}}{1440a^2\sin\theta} \left[1 + \left(\beta - \frac{3}{8}\right) \frac{a}{R_1+R_2} \right].
\end{equation}
For this particular geometry, the interaction energy has been computed numerically in \cite{Rodriguez}. The numerical results reproduce Equation \eqref{incl_cyl} at short distances.

The results obtained for the $\beta$-coefficients in each case are summarized in Table \ref{betas}.
\\

\begin{table}[t]
\begin{center}
\begin{tabular}{| c | c | c | c | c |} \hline 
$\beta_{\rm D}$  & \boldmath{$\beta_{\rm N}$} & \boldmath{$\beta_{\rm DN}$} & \boldmath{$\beta_{\rm ND}$} & \boldmath{$\beta_{\rm EM}$} \\ \hline
2/3 & $2/3(1 - 30/\pi^2)$ & 2/3 & $2/3 - 80/7\pi^2$ & $2/3(1 - 15/\pi^2)$ \\ \hline
\end{tabular}
\caption{$\beta$ coefficient {from} 
(\ref{bimonte}) for the following five cases: a scalar field obeying 
{Dirichlet (${\rm D}$)} or {Neumann ($\rm N$)} 
boundary conditions on both surfaces, or ${\rm D}$ boundary condition on one surface and ${\rm N}$ boundary condition on the other, or vice versa, and for the {electromagnetic (${\rm EM}$)} 
field with ideal metal boundary conditions~\cite{Bimonte1}}
\label{betas}
\end{center}
\end{table} 

Having presented in this Section a derivation and some interesting  results obtained by applying  the DE to 
the Casimir effect at zero temperature and for perfect boundary conditions, we present in the rest of the review 
some generalizations and applications.

\section{Finite Temperature, Nonanalyticities, and DE}
\label{sec:finite T}
The DE can be extended to the finite temperature case \cite{Bimonte2,FLMTdim,FLMTem}, the free energy being 
the relevant functional to approximate. 
There are at least two reasons why this  extension is not trivial:  firstly, the temperature introduces a 
dimensionful magnitude, and this will reflect itself in the form of the DE (part of it was fixed by dimensional
analysis).  Second, a known phenomenon  in quantum field theory at finite temperature is the so-called "dimensional reduction", by which a bosonic  model which is defined in $d+1$ dimensions at 
zero temperature, becomes effectively $d$-dimensional at  high temperatures.  The DE should 
therefore manifest (and interpolate between) those two cases.

We first {describe, in Section} {\ref{sec61}}, the results for a scalar field satisfying Dirichlet 
conditions~\cite{FLMTdim} in $d+1$ dimensions. {Then,} {Section \ref{sec62} discusses} 
the appearance of nonanaliticities for Neumann boundary conditions~\cite{FLMTdim,FLMNeu}. Finally, 
we comment on the results for the EM field with imperfect boundary conditions~\cite{FLMTem,Bimonte2} {(Section \ref{sec63})} {and on semianalytic formula for  plane--sphere geometry} {(Section \ref{sec64})}.

\subsection{Dirichlet Boundary Conditions}
\label{sec61}

In the finite-temperature case, and for the same geometry that we have considered in the zero temperature case, 
the functions $V(\psi)$ and $Z(\psi)$ cannot be completely determined from dimensional analysis alone. 
Indeed, on general grounds, we can assert that the Casimir free energy in $d+1$ dimensions, if the DE is applicable,
must have the form:
\begin{widetext}
\begin{equation}
	F_{\rm DE} [\psi] \;=\; \int d^{d-1}{\mathbf x}_\shortparallel \, 
\Big\{ b_0(\frac{\psi}{\beta},d) \frac{1}{[\psi({\mathbf x}_\shortparallel)]^d} 
\,+\, 
b_2(\frac{\psi}{\beta},d) \, \frac{(\nabla\psi)^2}{[\psi({\mathbf x}_\shortparallel)]^d} 
\Big\}  \label{DE dir}
\end{equation}\end{widetext}
where $b_0$ and $b_2$ are dimensionless and depends on the ratio of the local distance between 
surfaces $\psi$ and the inverse temperature $\beta$. They can be obtained from
the knowledge of the Casimir free energy for small departures around 
$\psi({\mathbf x_\shortparallel}) = a = {\rm constant}$. 
They are given by \cite{FLMTdim}
\begin{eqnarray}
 b_0(\xi)&=&\frac{\xi}{2}\sum_{n=-\infty}^{\infty}\int \frac{d^{d-1}\mathbf p_\shortparallel}{(2\pi)^{d-1}}\log\left(1-e^{-2\sqrt{(2\pi n\xi)^2+(\mathbf p_\shortparallel)^2}}\right) \nonumber \\
 b_2(\xi)&=& \frac{1}{2}\left [
\frac{\partial F^{(2)}(\xi,n,\mathbf l_\shortparallel)}{\partial|\mathbf l_\shortparallel|^2}\right]_{n\to 0,|\mathbf l_\shortparallel|\to 0}\,
\end{eqnarray}
where
\begin{widetext}
\begin{equation}
F^{(2)}(\xi; n, |{\mathbf l}_\shortparallel|) =  -2 \xi \, \sum_{m=-\infty}^{+\infty} 
\int \frac{d^{d-1}{\mathbf p}_\shortparallel}{(2\pi)^{d-1}} \Big\{
\frac{\sqrt{(2 \pi m \, \xi)^2 + {\mathbf p}^2_\shortparallel}}{1 - \exp\big[- 2 \sqrt{(2 \pi m \xi)^2  + {\mathbf p}^2_\shortparallel}\big]}  
\frac{\sqrt{(2 \pi (m +n) \, \xi)^2 + ({\mathbf p}_\shortparallel + {\mathbf l}_\shortparallel)^2}}{\exp\big\{2 \sqrt{ [2 \pi (m + n)  \xi]^2 + 
({\mathbf p}_\shortparallel + {\mathbf l}_\shortparallel)^2}\big\} - 1 }\Big\} \, .
\label{eq:fb3} \end{equation}
\end{widetext}
In the zero temperature limit, the Matsubara sum becomes an integral that can be analytically computed. The results are described in Table \ref{2}. The ratio $b_2/b_0$ tends to $1$ for large values of $d$.

\begin{table}[t]
\begin{center}
\begin{tabular}{| c | c | c |} \hline 
 {\bf Dimension} 
& {\boldmath{${b_2(d)}/{b_0(d)}$}} & \textbf{Approximate Value} \\ \hline
$d=1$ &$\frac{1}{\pi^2}\left(1+\frac{\pi^2}{3}\right)$ &$0.435$ \\ 
$d=2$ & $\frac{1+6\zeta(3)}{12\zeta(3)}$ &$ 0.569$   \\ 
$d=3$ & $2/3$ & $0.667$ \\ 
$d=4$ & $\frac{-\zeta (3)+10 \zeta (5)}{12 \zeta (5)}$ & $0.737$ \\ 
$d=5$ & $\frac{10 \pi ^2-21}{10 \pi ^2}$ & $0.787$\\ 
$d=6$ & $\frac{-2 \zeta (5)+7 \zeta (7)}{6 \zeta (7)}$ & $0.824$\\ \hline
\end{tabular}
\caption{Values of the ratio {${b_2(d)}/{b_0(d)}$} for different dimensions. The ratio tends to $1$ for $d\to\infty$.
See text for details.}
\label{2}
\end{center}
\end{table}

In the high temperature limit, we find
\begin{eqnarray}\label{eq:b0b2high}
&& \big[b_0(\xi,d)]_{\xi>>1} \;\simeq\; \xi \, \big[b_0(\xi, d-1)]_{\xi \to 0}\equiv \xi\,  b_0(d-1) \;, \nonumber\\
&& \big[b_2(\xi,d)\big]_{\xi>>1} \simeq \xi  \big[b_2(\xi, d-1)\big]_{\xi
\to 0}\equiv \xi  b_2(d-1)
\end{eqnarray} 
where $\xi = \psi/\beta$. The coefficients $b_0(d-1)$ and $b_2(d-1)$ agree with those for perfect mirrors 
at zero temperature, but in $d-1$ dimensions,  i.e., the `~dimensional reduction'' effect.

An interesting result is found when this is  applied to the (Dirichlet) Casimir interaction for a system consisting of 
a sphere in front of an infinite plane. Denoting by $a$ the distance
between the surfaces, and by $R$ the radius of the sphere, we get for the free energy at high temperatures:
\begin{equation}
F _{\rm DE} [\psi]\vert_{\psi/\beta >>1,d=3} \,\sim\, -\frac{\zeta(3)R}{8\beta a} \left(1-\frac{1}{6\zeta(3)} \frac{a}{R}\log\left(\frac{a}{R}\right)\right)\, .
\label{Fsp}
\end{equation} 
We see that the $R/a^2$-behavior corresponding to the dominant 
contribution at  zero temperature  changes to $R/a\beta$  in the  high temperature case.  This could be expected on dimensional grounds, if one assumes that the free energy is linear in the temperature in this limit.
Note that the same problem has been exactly solved in Ref.\cite{bimonteprl}, and one can  show that Equation  (\ref{Fsp}) does agree with the small-distance expansion obtained from the exact solution.

It is worth to remark that the  NTLO correction from the DE becomes nonanalytic,
because of the integration, in the ratio $a/R$.  This behavior has been observed in numerical
calculations of the Casimir interaction energy for this geometry, in the
infinite temperature limit, for the electromagnetic case (see
Refs.\cite{bimonteprl,Neto2012}). It is important to recognize that this nonanalyticity  has nothing to do with the
nonanalyticity in momenta of the form factors described in Section 4,  and is a non trivial 
prediction of the DE.

\subsection{Neumann Boundary Conditions}
\label{sec62}

This case, discussed in Ref.\cite{FLMNeu},  highlights a potential warning to the applicability of the DE, already mentioned previously:  the appearance of nonanalyticities in the form factors. To begin with,  
we deal with the zero temperature case in $2+1$ dimensions, since the nonanalyticity appears  because of the 
existence of a Matsubara mode which behaves as a massless field in $2+1$ dimensions, with Neumann boundary conditions. 

The free Euclidean action for the vacuum (i.e., $T=0$) field $\varphi$ is given by
\begin{equation}\label{eq:defsphi}
{\mathcal S}_0\;=\; \frac{1}{2}\,\int d^3x \, (\partial \varphi)^2\;,
\end{equation}
and, instead of imposing perfect Neumann boundary conditions on the surfaces, we add the following action to 
describe the interaction between the vacuum field and the mirrors:
\begin{widetext}
\begin{equation}\label{intscalar}
{\mathcal S}_{\rm I}[\varphi] = \frac{1}{2\bar\mu} \int d^3x \left[ 
\delta(x_2) (\partial_2\varphi(x))^2 
+ \sqrt{g(x_\parallel)} \delta(x_2 - \psi(x_\parallel)) (\partial_n
\varphi(x))^2 \right] \;.
\end{equation} \end{widetext}
The constant $\bar\mu$, which has the dimensions of a mass, is used to impose Neumann boundary conditions in 
the $\bar\mu\to 0$ limit. We use the same $\bar\mu$ on both L and R mirrors, since we will assume them to have 
identical properties, differing just in their position and geometry.

The DE approximation to the Casimir energy can be computed following standard steps. The result reads, in the limit $\bar\mu \psi \to 0$ \cite{FLMNeu}, 
\begin{widetext}
\begin{equation}\label{eq:almostneumanm}
E_{\rm DE}[\psi] \;=\; -\frac{1}{16 \pi}\int_{-\infty}^\infty dx_1 \, \frac {1}{\psi(x_1)^2}
\left[ \zeta(3)
+  \log[\bar\mu\psi(x_1)]   \left(\frac{d\psi(x_1)}{dx_1} \right)^2  \right] \;.
\end{equation} 
\end{widetext}
In the expression above, the first term is the PFA contribution while  the second one is a non trivial 
correction to it, and depends on the shape of the boundary (defined
by $\psi$). 
It is then clear that, as this equation shows, the DE is well posed when imposing imperfect Neumann boundary conditions in $2+1$ dimensions. On the contrary,  it cannot be  applied when the boundary conditions become perfect ($\bar\mu=0$).
The reason is that the hypothesis of analyticity in momentum, used to derive the DE, is clearly violated.
The non-existence of a local expansion is due to the existence of  massless modes,  allowed by Neumann 
boundary conditions.

Since, at finite temperatures,  a $3+1$ dimensional theory may be decomposed into the sum of an infinite tower 
of decoupled $2+1$ dimensional Matsubara modes, each one satisfying N boundary conditions, and with a mass $\frac{2n\pi}{\beta}$, $n = 0, 1, 2, \ldots$   The existence of the massless $n=0$ mode (the only one surviving in the high temperature limit)  means  that analyticity will be lost in $3+1$ dimensions,  for any non zero temperature. 
That  is indeed the case \cite{FLMTdim}. 
We summarize here some of the main features of that example: the free energy in the $d+1$ dimensional 
Neumann case can again  be written as before  (see Eq.(\ref{DE dir}), but with coefficients $c_0$ and $c_2$ instead of $b_0$ and $b_2$.
The zero order term coincides with the one for the Dirichlet case; namely: $c_0 = b_0$. 

When $d=3$, the NTLO term contains, besides a  local term, a  nonlocal contribution which is
linear in $T$, and thus present for any $T>0$.  Hence, there is no  local DE  for perfect 
Neumann boundary conditions at $d=2$ at zero temperature and for $d=3$ at any finite temperature.
Indeed, an expansion for small values of $\vert k_\shortparallel\vert$ of the form factor contains, in addition to  a
term proportional to $k_\shortparallel^2$, one proportional to $(Ta) k_\shortparallel^2
\log(k_\shortparallel^2a^2)$.

\subsection {The Electromagnetic Case  for Imperfect Boundary Conditions}
\label{sec63}

We have seen that, for  a real scalar field in the presence of Neumann boundary conditions,
the DE cannot be applied when in  $2+1$ dimensions  at zero temperature, or in $3+1$ dimensions at a non-zero  temperature~\cite{FLMTdim}.   The reason is that, as we have shown,  nonanalyticities in the form factors appear.
We have shown that  the nonanalyticity could be cured  by introducing a small departure from perfect Neumann conditions~\cite{FLMNeu}.  It is natural to wonder whether  the  nonanalyticities could also be cured by a 
similar approach for the EM field in $3+1$ dimensions at finite temperatures.
We know,  based on the insight obtained from Ref.\cite{FLMNeu},  that nonanalyticities are originated in contributions due to  dimensionally reduced massless modes:  zero Matsubara frequency terms. 
To obtain an answer to this question, in Ref.\cite{FLMTem} we singled out  in detail  the zero-mode contributions 
to the free energy, for a media described by non trivial permittivity $\epsilon(\omega)$ and permeability $\mu(\omega)$ functions.

 We start from the free energy $F$ for the EM field, which  can be written in terms of the partition 
 function ${\mathcal Z}(\psi)$,  as follows: 
\begin{equation}
	F(\psi)\,=\,-\frac{1}{\beta}\,\log\big[\frac{{\mathcal
	Z}(\psi)}{{\mathcal Z}_0}\big]\;,
\end{equation}
where the denominator, ${\mathcal Z}_0$, denotes the partition function for the EM field in the absence of media and 
\begin{equation}\label{eq:defzbeta}
{\mathcal Z}(\psi) \;=\; \int \big[{\mathcal D}A\big] \;
e^{-{\mathcal S}_{\rm inv}(A)}\, .
\end{equation} 
The gauge invariant action ${\mathcal S}_{\rm inv}(A)$ reads \begin{widetext}
\begin{equation}\label{eq:defsinv}
{\mathcal S}_{\rm inv}(A)\;=\;\int_0^\beta  d\tau \int_0^\beta d\tau' \int d^3{\mathbf x} \,
\big[  \frac{1}{2} \, F_{0j}(\tau,{\mathbf x})
\epsilon(\tau-\tau', {\mathbf x}) F_{0j}(\tau',{\mathbf x}) 
 +  \frac{1}{4} F_{ij}(\tau,{\mathbf x})
\mu^{-1}(\tau-\tau', {\mathbf x})  F_{ij} (\tau,{\mathbf x}) \big] \;.
\end{equation}\end{widetext}
Here, indices like $i,\,j\, \ldots$  run over spatial indices, Einstein  summation convention is assumed, 
and $\epsilon(\tau-\tau' , {\mathbf x})$ and $\mu(\tau-\tau' , {\mathbf x})$  denote the imaginary time versions of the permittivity and permeability, respectively ($\mu^{-1}$ is the inverse integral kernel of $\mu$).

The geometry of the system is determined by same two surfaces $L$ and $R$ we have considered before, and 
defined by $x_3=0$ and $x_3=\psi(\mathbf x_\shortparallel)$, but now they correspond to the boundaries of the media,
i.e., 
\begin{widetext}
\begin{eqnarray}
\epsilon(\tau-\tau', {\mathbf x}) &=& 
\theta(-x_3) \epsilon_L(\tau-\tau') +
\theta(x_3) \theta(\psi({\mathbf x}_\shortparallel) - x_3)+
\theta(x_3-\psi({\mathbf x}_\shortparallel)) 
\epsilon_R(\tau-\tau') \nonumber \\
\mu(\tau-\tau', {\mathbf x}) &=& 
\theta(-x_3) \mu_L(\tau-\tau') +
\theta(x_3) \theta(\psi({\mathbf x}_\shortparallel) - x_3)+
\theta(x_3-\psi({\mathbf x}_\shortparallel)) \,\mu_R(\tau-\tau'),\nonumber 
\end{eqnarray}
\end{widetext}
where $\epsilon_{L,R}(\tau-\tau')$ and $\mu_{L,R}(\tau-\tau')$ characterize the permittivity and permeability of the respective mirror. 

We can expand the fields and the electromagnetic properties  as
\begin{eqnarray}\label{eq:fou1}
A_\mu (\tau,{\mathbf x}) &=& \frac{1}{\beta} \,
\sum_{n=-\infty}^{+\infty} \widetilde{A}_\mu^{(n)}({\mathbf x}) \, e^{i \omega_n
\tau} \nonumber\\
\epsilon(\tau-\tau',{\mathbf x}) &=& \frac{1}{\beta} \,
\sum_{n=-\infty}^{+\infty} \widetilde{\epsilon}^{(n)}({\mathbf x}) \, e^{i \omega_n
(\tau-\tau')} \nonumber\\
\mu(\tau-\tau',{\mathbf x}) &=& \frac{1}{\beta} \,
\sum_{n=-\infty}^{+\infty} \widetilde{\mu}^{(n)}({\mathbf x}) \, e^{i \omega_n
(\tau-\tau')} 
\end{eqnarray}   
where $\omega_n \equiv 2\pi n/\beta$ ($n \in {\mathbb Z}$) are the Matsubara frequencies. 

Inserting these expansions into the partition function one can readily check the factorization
\begin{equation}\label{eq:zprod}
{\mathcal Z}(\psi) \;=\; \prod_{n=0}^\infty {\mathcal Z}^{(n)}(\psi) \; ,
\end{equation}
and therefore
\begin{equation}
F(\psi)\;=\; \sum_{n=0}^\infty \, F^{(n)}(\psi) \;.
\end{equation}

As mentioned, we are particularly interested in the $n=0$ contribution, 
\begin{equation}
{\mathcal Z}^{(0)}(\psi) \;=\; \int [{\mathcal
D}\widetilde{A}_0^{(0)}{\mathcal D}\widetilde{A}_j^{(0)}]\;e^{- {\mathcal
S}^{(0)}(\widetilde{A}_0^{(0)}, \widetilde{A}_j^{(0)})}\, ,
\end{equation}
where
\begin{widetext}
\begin{equation}
{\mathcal S}^{(0)}(\widetilde{A}_0^{(0)}, \widetilde{A}_j^{(0)}) \,=\, 
\frac{1}{\beta}
\int d^3{\mathbf x} \,
\big[ \frac{1}{2} \, \widetilde{\epsilon}^{(0)}({\mathbf x}) 
(\partial_j \widetilde{A}_0^{(0)})^2 
\,+\, 
\frac{1}{4 \,\widetilde{\mu}^{(0)}({\mathbf x})} (\widetilde{F}_{jk}^{(0)})^2 
\,+\, \frac{1}{2} \, \Omega_0^2({\mathbf x}) (\widetilde{A}_j^{(0)})^2
\big]\, ,
\end{equation}
\end{widetext}
and
\begin{equation}
\Omega_0^2({\mathbf x}) \,\equiv\, \lim_{n\to 0} \,
\big[ \omega_n^2 \,\widetilde{\epsilon}^{(n)}({\mathbf x}) \big]\;.
\end{equation}
Note that $\Omega_0$ vanishes for a dielectric and also for a metal described by the Drude model. 
On the other hand, it equals  the plasma frequency for a metal described by the plasma model.  

The zero mode contribution to the free energy therefore splits into a scalar ($s$) and a vector ($v$) contribution, the former associated to the field $\widetilde{A}_0^{(0)}$ and the later to $\widetilde{A}_j^{(0)}$
\begin{equation}
	F^{(0)} \;=\; F_s(\psi) \,+\, F_v(\psi) \;.
\end{equation}

To discuss the emergence of non-analyticities in the derivative expansion we computed $F_s$ and $F_v$
assuming $\psi(\mathbf x_\shortparallel)=a+\eta(\mathbf x_\shortparallel)$ up to second order in $\eta$. The quadratic contributions can be written as
\begin{equation}
	F_{s,v}^{(2)}\,=\, \frac{1}{2} \int \frac{d^2k_\parallel}{(2\pi)^2}
	\,f_{s,v}^{(2)}(k_\shortparallel, a) \, |\tilde{\eta}(k_\shortparallel)|^2 \;,
\end{equation}
and the crucial point is whether the functions  $f_{s,v}^{(2)}$are analytic or not in $k_\parallel$. 

Omitting the details, we summarize the main results \cite{FLMTem}:
for  finite values of $\mu$ and $\epsilon$, the scalar contribution $f_{s}^{(2)}$ analytic, including the limit $\epsilon\to \infty$, in which it tends to the $2+1$ dimensional Dirichlet value.  It develops a nonanalytic (logarithmic) contribution for $\mu=\infty$, since the kernel corresponds in this case to that of a scalar field in $2+1$ dimensions satisfying Neumann boundary conditions. In other words, magnetic materials regulate the non-analyticity of the TE zero mode.

On the other hand, the TM zero mode is nonanalytic whenever $\omega^2\epsilon(\omega)\to \Omega^2\neq 0$ as $\omega\to 0$  for both
mirrors.  
In terms of the models usually considered in the Casimir literature to describe real
materials, this condition  corresponds to the  plasma model. 

In summary, the nonanalyticities we observed for perfect conductors in our
previous work \cite{FLMTdim}, survive only under the assumption of perfectly
lossless materials. The NTLO corrections to PFA for metals (gold) at room temperature have been computed in Ref. \cite{Bimonte2}.

\subsection{A Semianalytic Formula for Plane-Sphere Geometry }
\label{sec64}

As a final application of the DE to compute the Casimir free energy
we mention the results of Ref.\cite{Bimonte3}, where 
 the author combined exact calculations for the zero mode and the DE to obtain a precise formula for the interaction between a sphere and a plane at a finite temperature which is valid at all separations.  We briefly describe here these findings.

Formally, the free energy for this geometry can be written as
\begin{equation}
 F = k_B T \sum_{n\geq 0}'Tr\log[1 -\hat M(i\xi_n)]\, ,    
\end{equation}
where the sum is over the Matsubara frequencies  $\xi_n=2\pi n k_BT/\hbar$ and $\hat M$ denotes scattering  matrix elements for this geometry. The prime on the sum indicates that the $n=0$ term has an additional $1/2$ factor. 

The $n=0$ contribution can be computed exactly using the Drude model to describe the materials of the plane and the sphere, and plays a crucial role.
Indeed, the proposed approximation for the Casimir force on the sphere of radius $R$ at a distance $a$ from the plane is 
\begin{equation}
F_{\rm approx}=F_{n=0}^{(\rm exact)}+ F_{n>0}^{(\rm PFA)}(1-\theta \frac{a}{R})\, ,
\end{equation}
where $\theta$ can be computed using the DE. Notably, $F_{\rm approx}$ describes with high precision the Casimir force at all separations, as can be checked by comparison with high precision numerical simulations of the exact scattering formula.

These results have been generalized in subsequent studies to the case of the two spheres- geometry \cite{Bimonte4},  also considering the differences that come from the use of the Drude vs plasma models, as well as for grounded vs isolated spheres \cite{Bimonte5}. The relevance of the use of grounded conductors in Casimir experiments has also been discussed in Ref. \cite{FLMgrounded}.


\section{Casimir-Polder Forces}\label{sec:CP}

The DE approach has also been applied to the calculation of the Casimir-Polder interaction between a polarizable particle and a gently curved surface \cite{BimonteCP}. We present in this Section a simplified version of the results contained in that reference.

When a small polarizable particle is at a distance $a$ of a planar surface, the Casimir-Polder potential reads \cite{Bordagbook}
\begin{equation}
U(a)=-\frac{1}{a^4}
\int\frac{d\xi}{2\pi}\alpha(i\omega_c\xi)\beta^{(0)}(\xi)\, ,
\end{equation}
where  $\alpha(\omega)$ is the frequency dependent polarizability (which is assumed isotropic), $\omega_c=c/a$,  and
\begin{equation}
\beta^{(0)}=\frac{e^{-2\xi}}{2}(1+ 2\xi+2\xi^2)\, .
\end{equation}
For moderate distances such that $\alpha(\omega)\approx \alpha(0)$ one obtains the usual Casimir-Polder potential \cite{Casimir-Polder}
\begin{equation}
U(a)=-\frac{3}{8\pi}\frac{\alpha(0)}{a^4}\, .
\end{equation}

Assume now that the particle is in front of a slightly curved surface. The particle is at the origin of coordinates, and the surface is described, as usual, by the height function $z=\psi(\mathbf x_\shortparallel)$. 
The DE for the Casimir-Polder interaction $U_{\rm DE}$ assumes that the interaction depends on the derivatives of the height function $\psi$ evaluated at $\mathbf x_\shortparallel=0$, the point on the surface closest to the particle (a local minimum for $\psi$).  If the surface is homogeneous and isotropic, then the interaction
energy must be invariant under rotations of the $\mathbf x_\shortparallel$ coordinates. The more general expression compatible with this properties describes the Casimir -Polder interaction energy at $T=0$ reads \cite{BimonteCP}:
\begin{equation}
U_{\rm DE}=
-\frac{1}{\psi^4} \int\frac{d\xi}{2\pi}\alpha(i\omega_c\xi)\left(\beta^{(0)}(\xi)+\beta^{(1)}(\xi)\psi\nabla^2\psi\right)\, .
\end{equation}

The dimensionless function $\beta^{(1)}$ can be read  from  
the perturbative expansion of the potential $U$, carried to second order in the deformation, that is, 
for
$\psi(\mathbf x_\shortparallel)= a + \eta({\mathbf x_\shortparallel})$ with $\eta({\mathbf x_\shortparallel})\ll a$.
We stress that here the Casimir-Polder energy is not a functional but a function of $\psi$ and its derivatives evaluated at the origin of coordinates (recall that $\nabla\psi(\mathbf 0)=0$).
The DE is expected to be valid when $a\ll R_1, R_2$, the radii of curvature of the surface at $\mathbf x_\shortparallel=0$. Note that $\psi(\mathbf 0)=d$ and
\begin{equation}
\nabla^2\psi (\mathbf 0)=\frac{1}{R_1}+\frac{1}{R_2}\, .
\end{equation}
Using again the static polarizability approximation, $\alpha(\omega)\approx \alpha(0)$,  one obtains
\begin{equation}
U_{\rm DE}=-\frac{1}{\pi a^4}\alpha(0)\left
[\frac{3}{8}-\frac{13}{60}a(\frac{1}{R_1}+\frac{1}{R_2})\right]\, .
\end{equation}

The results presented in Ref. \cite{BimonteCP} are much more general than those described here: they include the Casimir-Polder potential for a general polarization tensor $\alpha_{\mu\nu}(\omega)$ and higher order corrections proportional to $(a/R_i)^2$, as well as the details of the computation of the corresponding functions $\beta^{(p)}$. Additional applications can be found in \cite{moreCP1,moreCP2}.

\section{Other Techniques Beyond PFA}
\label{sec:other}

In Ref. \cite{pauloEPL} a detailed analysis of the Casimir effect’s roughness correction in a setting involving parallel metallic plates is presented. The plates were defined through the plasma model. The approach used is perturbative, factoring in the roughness amplitude and allowing for the consideration of diverse values of the plasma wavelength, plate separation, and roughness correlation length. A notable finding was that the roughness correction {\em exceed\/} the predictions of the PFA. The authors have calculated
the second-order response function, $G(k)$, across a spectrum of values encompassing the plasma wavelength 
($\lambda_P$), distance ($a$), and roughness wave vector ($k$):
\begin{widetext}
\begin{equation}
    G(k) = -\frac{A}{8\pi^2}\frac{1}{a^5q}\int_0^\infty \frac{dKe^{-2K}}{1 - e^{-2K}} \int_{\vert K-q\vert}^{K+q} dK' \frac{(KK')^2 + \frac{1}{4} (K^2 + K'^2-q^2)^2}{1 - e^{-2K'}} \,,
\end{equation}
\end{widetext}
applicable when \ $\lambda_p \rightarrow 0$. Here, $A$ represents the plate surface area, $K$ the dimensionless 
integration variable denoting the imaginary wave vector's $z$-component scaled by plate separation $d$, $K'$
the longitudinal component of the imaginary wave vector for the diffracted wave, and $q = ka$. 
 
The calculation in Ref.~\cite{pauloEPL} helps to compute the second-order roughness correction as a function of the surface profiles, $h_1$ and $h_2$.
Analytical solutions were determined for specific limiting cases, revealing a more complex relationship with the perfect reflectors
model than previously recognized~\cite{pauloEPL03,emigPRL01}, particularly in scenarios involving extended distances and small
roughness wavelengths. While the asymptotic case of long roughness wavelengths aligns with PFA predictions, it was established 
that PFA generally underestimates the roughness correction, a critical aspect for exploring constraints on potentially new weak forces at 
sub-millimeter ranges.

As a further expansion to~\cite{pauloEPL}, in Ref. \cite{pauloUniverse21}, the authors explored the Casimir interaction 
between a plane and a sphere of radius $R$ at a finite temperature $T$, in terms of  the distance of closest approach, $a$.  
Noting that, under the  usual experimental conditions, the thermal wavelength $\lambda_T$ satisfies $a\ll \lambda_T \ll R$, 
they evaluated the leading correction to the PFA,  applicable to such intermediate temperatures.  
They resorted to developing the scattering formula in the plane-wave basis.  The result captures the combined effect of 
spherical geometry and temperature, and is expressed as a sum of temperature-dependent logarithmic terms. 
Remarkably, two of these logarithmic terms originated from the Matsubara zero-frequency contribution. 

Defining the variables $x = a/R$ and $\tau = a/\lambda_T$, and the deviation $\delta {F}(T) = {F}(T) - {F}(0)$, in the intermediate temperature regime $x \ll \tau \ll 1$,  it is found in Ref. \cite{pauloUniverse21} that
\begin{widetext}
\begin{equation}
    \Delta  =  \frac{\delta {F}(T) - \delta {F}_{\rm PFA}(T)}{{F}_{\rm PFA}(T)} 
     \approx  \frac{45}{\pi^3} x\tau \left[- \log^2(x)+ 2 [1 - \log(2)]\log(x)+2\log^2(\tau) + {\cal O}(\log(\tau))\right].
\end{equation}
\end{widetext}
The leading neglected terms stem from non-zero Matsubara frequencies. 

In Ref. \cite{pauloOSA}, the leading-order correction to PFA in a plane-sphere geometry was derived. 
The momentum representation connected this with geometrical optics and semiclassical Mie scattering. The primary contributions 
are shown to come from diffraction, with TE polarization becoming more relevant than TM polarization. The diffraction contribution is
calculated at leading order, using the saddle-point approximation, considering leading order curvature effects at the 
sphere tangent plane. 

Additionally, the next-to-leading order (NTLO) term in the saddle-point expansion contributed to the PFA correction. 
This involved computing the round-trip operator within the WKB approximation, representing sequences of reflections 
between the plane and the sphere. A key aspect was the tilt in the scattering planes, allowing TE and TM 
polarizations to mix.

Comprehending the implications of polarization mixing channels on the geometric optical correction applied to PFA holds considerable importance. Indeed, these channels are recognized for inducing negative Casimir entropies with a geometric foundation.~\cite{24–26OSA1,24-26OSA2,24-26OSA3,49-51OSA1,49-51OSA2,49-51OSA3}. In spite of the non-vanishing 
contribution of the polarization mixing matrix elements, the total correction associated with the tilt 
between the scattering and Fresnel planes is zero at NTLO. This implies that 
the primary  correction to the PFA would remain unchanged even if the complexities arising from the differences 
between the Fresnel and scattering polarization bases were initially ignored. 
The latter points to the fact that a different approach, one that completely omits the effect of polarization mixing, 
could directly produce the leading order correction to PFA. Plane waves proved to be a well-suited basis for 
studying the Casimir effect, as has been evidenced in the more recent study ~\cite{paulo22}. The utility of that basis ranges from 
analytical to numerical applications, particularly when dealing with objects in close proximity, the most relevant
situation in experiments. 
It has been also shown that the use of plane waves was notably effective in improving the interpretation of results 
in the realms of geometrical optics and diffractive corrections.

In the context of a setup involving two spheres with arbitrary radii in vacuum, it was shown in \cite{paulo22} that the PFA  emerged as the 
leading term in an asymptotic expansion for large radii. Extending a prior calculation based on the saddle-point approximation,
involving a trace over multiple round-trips of electromagnetic waves between the spheres, the study encompassed 
spheres made of bi-isotropic material, requiring  the consideration of polarization mixing during reflection processes. 
The result was naturally elucidated within the framework of geometrical optics. 

Then, by relying on  a saddle-point approximation framework, the authors derived leading-order corrections,  
of geometrical and diffractive origins. Explicit results, at first obtained for perfect electromagnetic 
conductors (PEMC) spheres at zero temperature, indicated that for certain material parameters, the PFA contribution vanishes;
should that be the case, the leading-order correction would be the dominant term in the Casimir energy.

In the lowest-order saddle-point approximation, but including diffractive corrections, one can show that 
the expression for the Casimir energy becomes:
\begin{equation}
    E_{\rm LO-SPA} = -\frac{\pi^3R_{\rm eff}}{720a^2} \left[ 1 - \frac{15}{\pi^2} x + \frac{15(10 + 3\pi)}{4\pi^3} x^{3/2} + ..\right], 
\end{equation}
where $x = a/R_{\rm eff}$. As expected, this result reproduces the PFA result and its leading-order diffractive correction. 
The NTLO correction behaves as $x^{3/2}$. However, the prefactor obtained accounts  for  about $90\%$ of the one coming 
from numerical results~\cite{pauloOSA}. This discrepancy may be traced back to having neglected the NTLO-SPA and 
NNTLO-SPA contributions.

\section{Conclusions}\label{sec:conc}
In this review, we have  discussed several properties and applications of the DE approach, mostly as a method to improve the predictions of the Proximity Force 
Approximation, of long standing use in many different fields.

We started the review by  briefly discussing the precursor of the PFA: the Derjaguin (and related)  approximations, since  we have  found them rather appropriate in order to display the essentially geometric nature of the kind of 
problem we discuss: two quite close smooth surfaces, and an interaction energy  between
them. Depending on the kind of system being considered, that interaction between the two surfaces 
may or not be the result of  the superposition of  the interactions between pairs.  An example of  an
interaction which is not the result of such a superposition is the Casimir effect. Note, however,  that even
when the fundamental interaction satisfies a superposition principle, like in electrostatics, the actual evaluation of the Coulomb integral to calculate the total interaction energy could be a rather involved problem because the actual charge density may not be known a priori. That is indeed the
case when the surfaces involved are conductors, since that usually requires finding  
the electrostatic potential.  We have used precisely this problem in order to present the idea of the 
DE in a concrete example:  to calculate the electrostatic energy between two conducting surfaces held 
at different potentials.  

After introducing and applying the DE in that example, we have discussed its more general proof of
that expansion, by first putting the problem in a more general and abstract way: how to approximate, 
under certain  smoothness assumptions, a functional of a pair of surfaces. At the same time, the proof provides a concrete way to determine the PFA and its NTLO correction, the DE: one just needs to perform an
expansion in powers of the deformation of the surfaces about the situation of two flat and  parallel surfaces. 

The derivations and examples here have been presented for a geometrical setting were one surface is a plane,
while the other may be described by a single Monge patch based on that plane.  However, as shown by
other authors, under quite reasonable and general assumptions,  the results obtained for that situation may be  generalized to the case of two curved surfaces parametrized by their respective patches, based on a common plane (which now does not coincide with one of the physical surfaces). 

Then  we  reviewed different applications of the DE to the zero temperature 
Casimir effect,  considering different  fields and boundary conditions, staring from the cases of  the scalar field with Dirichlet boundary conditions,  then the EM  field in the presence of perfectly conducting surfaces, and commented on the scalar field with Neumann conditions.

We afterwards presented a description and brief review of  the extension of DE to finite temperature cases, and different
numbers of spatial dimensions. The temperature is  a dimensionful magnitude  and the 
phenomenon of dimensional reduction presents a problem when there are Neumann boundary conditions or 
when an EM field is involved. Indeed, dimensional reduction implies the existence of a massless 
$2+1$ dimensional field (with Neumann conditions), and this mode introduces a nonanalyticity  in momentum
space,  which violated one of the hypothesis of the DE, and therefore it cannot be applied.  Nevertheless, 
we have shown that the  introduction of a small departure from ideal Neumann conditions solves this
issue, namely, analyticity is recovered and the DE may be applied.

We also mentioned the application of DE to the Casimir-Polder interaction, particularly between a polarizable particle and a gently curved surface. This example  highlights the broader implications of DE in understanding particle-surface interactions beyond the Casimir force itself.

To conclude, we have presented in this review the main features of the DE approach, with a focus in the 
Casimir effect, but pointing at the fact that its applicability can certainly go beyond that realm. We have 
shown that explicitly for electrostatics, but we expect it to be applicable to, for example, the same kind of
systems where the DA, SEI and SIA were introduced.

\vspace{6pt}

This research was funded by Agencia Nacional de Promoci\'on Cient\'{\i}fica y Tecnol\'ogica
(ANPCyT), Consejo Nacional de Investigaciones Cient\'{\i}ficas y T\'ecnicas (CONICET), Universidad de Buenos Aires (UBA),
 and Universidad Nacional de Cuyo (UNCuyo), Argentina. 

\end{document}